\documentclass[journal]{IEEEtran}
\usepackage{tensor}
\usepackage{graphicx}
\usepackage{subcaption}
\usepackage{tabularx}
\usepackage[cmex10]{amsmath}
\usepackage{amsthm}
\usepackage{amssymb}
\usepackage{bm}
\usepackage{algorithm}
\usepackage{algorithmic}
\usepackage{cite}
\usepackage{setspace}
\usepackage{stfloats}
\usepackage{enumerate}
\usepackage{cases}
\usepackage{multirow}
\usepackage{mathrsfs}
\usepackage{array}
\usepackage{color}
\usepackage{verbatim}
\usepackage{extpfeil}
\usepackage{hyperref}
\usepackage{booktabs}
\usepackage{multirow}
\usepackage{caption}
\usepackage{graphicx}
\usepackage{tabularx}
\usepackage{epstopdf}
\usepackage{textcomp}
\usepackage{epsfig,epsf,color,balance,cite}
\usepackage{amssymb}
\usepackage{amsthm}
\usepackage{graphicx}
\usepackage{array,color}
\usepackage{amsmath}
\usepackage{graphicx}
\usepackage{tabularx}
\usepackage{epsfig,epsf,color,balance,cite}
\usepackage{algorithmic}
\usepackage{algorithm}
\usepackage{bm}
\usepackage{caption}
\usepackage{textcomp}
\usepackage{color}
\usepackage{multirow}
\usepackage{cite}
\usepackage{enumerate}
\usepackage{cases}
\usepackage{color}
\usepackage{epstopdf}
\usepackage{textcomp}
\usepackage{url}
\newtheorem{remark}{Remark}
\def\BibTeX{{\rm B\kern-.05em{\sc i\kern-.025em b}\kern-.08em
    T\kern-.1667em\lower.7ex\hbox{E}\kern-.125emX}}
\begin{document}

	\makeatletter
	\newcommand{\rmnum}[1]{\romannumeral #1}
	\newcommand{\Rmnum}[1]{\expandafter \@slowromancap \romannumeral #1@}
	\makeatother
	
	\title{Exploit High-Dimensional RIS Information to Localization: What Is the Impact of  Faulty Element? }
	
	\author{ Tuo Wu, Cunhua Pan, \emph{Senior Member, IEEE}, Kangda Zhi,  Hong Ren, \emph{Member, IEEE},  Maged Elkashlan, \emph{Senior Member, IEEE}, Cheng-Xiang Wang, \emph{Fellow, IEEE}, Robert Schober,  \emph{Fellow, IEEE}, Xiaohu You, \emph{Fellow, IEEE}
\thanks{\emph{(Corresponding author: Cunhua Pan.)}}
\thanks{ The work of  Tuo Wu was supported by the China Scholarship Council (CSC).  The work of Cunhua Pan was supported in part by the National Natural Science Foundation of China (Grant No. 62201137,62331023). The work of Cunhua Pan and Hong Ren was supported in part by the Fundamental Research Funds for the Central Universities, 2242022k60001, and by the Research Fund of National Mobile Communications Research Laboratory, Southeast University (No.2023A03). The work of Hong Ren was supported in part by the National Natural Science Foundation of China (Grant No. 62101128, 62350710796). The work of Cheng-Xiang Wang was supported by the National Natural Science Foundation of China (NSFC) under Grant 61960206006, the Fundamental Research Funds for the Central Universities under Grant 2242022k60006, the Key Technologies R$\&$D Program of Jiangsu (Prospective and Key Technologies for Industry) under Grants BE2022067 and BE2022067-1, and the EU H2020 RISE TESTBED2 project under Grant 872172. The work of Robert Schober was partly supported by the Federal Ministry of Education and Research of Germany under the programme of ``Souveran. Digital. Vernetzt.'' joint project 6G-RIC (project identification number: PIN 16KISK023) and by the Deutsche Forschungsgemeinschaft (DFG, German Research Foundation) under grant SCHO 831/15-1.}
\thanks{T. Wu  and M. Elkashlan are with the School of Electronic Engineering and Computer Science at Queen
Mary University of London, London E1 4NS, U.K. (Email:\{tuo.wu,  maged.elkashlan\}@qmul.ac.uk). C. Pan, H. Ren, C.-X. Wang, and X. You are with the National Mobile Communications Research Laboratory, Southeast University, Nanjing 210096, China. (e-mail: \{cpan, hren, chxwang, xhyu\}@seu.edu.cn). K. Zhi is with the School of Electrical Engineering and Computer Science, Technical University of Berlin, 10623 Berlin (e-mail: k.zhi@tu-berlin.de).   R. Schober is with the Institute for Digital Communications, Friedrich Alexander-University Erlangen-N$\ddot{\textrm u}$rnberg (FAU), Germany.  (e-mail: robert.schober@fau.de).}
}
	
	\markboth{}
	{}
	\maketitle

	\begin{abstract}
		This paper proposes a novel localization algorithm using the reconfigurable intelligent surface (RIS)  received signal, i.e., RIS information. Compared with  BS received signal, i.e., BS information, RIS information offers  higher dimension and richer feature set, thereby providing an enhanced capacity to  distinguish positions of the mobile users (MUs). Additionally, we address a practical scenario where RIS contains some unknown (number and places) faulty elements that cannot receive signals.  Initially, we employ transfer learning to design a two-phase transfer learning (TPTL) algorithm, designed for accurate detection of faulty elements. Then our objective is to regain the information lost from the faulty elements and reconstruct the complete high-dimensional RIS information for localization. To this end, we propose a transfer-enhanced dual-stage (TEDS) algorithm. In \emph{Stage I}, we integrate the CNN and variational autoencoder (VAE) to obtain the RIS information, which in \emph{Stage II},  is input to the transferred DenseNet 121 to estimate the location of the MU. To gain more insight, we propose an alternative algorithm named transfer-enhanced direct fingerprint (TEDF) algorithm which only requires the BS information. The comparison between TEDS and TEDF reveals the effectiveness of faulty element detection and the benefits of utilizing the high-dimensional RIS information for localization. Besides, our empirical results demonstrate that the performance of the localization algorithm is dominated by the high-dimensional RIS information and is robust to unoptimized phase shifts and signal-to-noise ratio (SNR).
	\end{abstract}

	\IEEEpeerreviewmaketitle

\section{Introduction}
The upcoming introduction of sixth generation (6G) Internet of Things (IoT) wireless networks marks a pivotal moment in our technological journey \cite{Wang1, ChenHui1, Wang3}. Enhanced localization accuracy becomes central in this new era  \cite{ChenHui2}.  The heightened demand for accuracy challenges us to not only innovate but to fundamentally rethink our approach to localization \cite{ Kaitao, HenkWymeersch1, HeJiguang1, Wang4}.  Recently, reconfigurable intelligent surfaces (RIS) \cite{Wang2, Cunhua1, Zhi, Gui1, Zhi3} have been introduced as a revolutionary technology to enhance the localization accuracy \cite{Wu1,JiguangHe2, Wang5} with several advantages. First, RIS offers a cost-effective and energy-efficient alternative to traditional base stations (BS) for maintaining consistent connectivity, especially in challenging environments. Additionally, their slim and adaptable design easily deploys in the urban structures, making them ideal for 6G networks.

{RIS-aided localization algorithms can be categorized into two main types:  \emph{two-step method} \cite{Alouini, chen2023multirisenabled} and \emph{fingerprint-based} \emph{method} \cite{WuFingerprint}. The \emph{two-step method} estimates parameters like angle of arrival (AoA) and time  of arrival (ToA) to deduce the location by exploiting geometric relationships \cite{ye2017power, hu2020deep, huang2018deep, yang2019deep}. The \emph{fingerprint-based method} \cite{ Bhattacherjee, 3DCNN}, on the other hand, relies on a database of pre-recorded signal characteristics, such as received signal strength (RSS), and compares real-time signals to the database for pinpointing a location. Building upon these two methods, algorithm design for RIS-aided localization has been widely investigated,  focusing on various applications and different perspective, such as multi-user \cite{Kamran2}, multi-RIS \cite{Wu2RIS}  near-field  \cite{NF1 ,NF2, Shubo,Xing,Mingan,Chongwen1,Cuneyd,Reza}, hardware impairment \cite{HW1, HW2, HW3},  and sidelink \cite{chen2023multirisenabled}.}

While the RIS-aided localization algorithm has been well-studied, however, the passive nature of the RIS has led to the design of these algorithms primarily based on the received signal at the BS, i.e., BS information \cite{ChenHui1}. However, due to the high cost and power consumption of radio-frequency chains may result in a limited number of antennas at the BS and therefore the dimension of received signal at the BS is low. This limitation could hinder the capacity to distinguish locations. Fortunately, an important but widely ignored fact is that the received signal at the RIS, i.e., RIS information, actually contains  much richer feature information about the user location than the BS signal which can be exploited to realize higher localization accuracy. This is because the RIS is commonly comprised of a large number of reflecting elements and therefore the dimension of its received signal could be  high enough, storing  sufficient  feature information, e.g., AoAs and ToAs \cite{WuFingerprint}, related to  user locations. By exploiting this high-dimensional RIS information, we can open a new  algorithm design degree-of-freedom (DoF) for RIS-aided localization.

While we recognize the significant potential, the localization design based on RIS information is indeed a highly challenging task  \cite{Cunhua1}. This challenge arises from the complexity of reconstructing high-dimensional RIS information. On the one hand, the passive nature of RIS poses a challenge in actively acquiring the signals received at the RIS. On the other hand, when employing the low-dimensional BS information to obtain the high-dimensional RIS information, the result is not unique. This is because the same BS information corresponds to multiple distinct RIS information. The uncertainty in the signal reconstruction process requires innovative algorithms to accurately capture  the RIS information.

{Let us consider a more general scenario where some of the RIS elements  may have been damaged (refer to as faulty elements in this paper) due to the environments conditions  \cite{HW1, HW2, HW3}.} In such case, localization algorithms designed under the assumption of a perfect RIS without faulty elements will no longer be reliable and may result in low localization accuracy. Furthermore,  the RIS faulty elements can reduce the dimension of the RIS received signal, leading to the localization information lost. This, in turn, significantly degrades the performance of the localization algorithm based on RIS information.

To mitigate the harmful impact of the faulty elements and regain the lost localization accuracy, it is crucial to first identify the faulty elements. Detecting faulty elements is equivalent to determining the status of each element of the RIS, classifying them as either faulty or non-faulty. This task, when applied to all  elements of the RIS, becomes a multi-label binary classification problem.  However, due to the passive nature of the RISs, conventional faulty detection algorithms designed for antenna are not capable of determining the number and the places of the faulty elements.

Given the presence of the faulty elements, the localization algorithm design based on the RIS information remains unexplored. To the best of the authors' knowledge, only   \cite{ozturk2023risaided}  has addressed the scenario in which the RIS contains faulty elements. However, the authors of \cite{ozturk2023risaided} designed the corresponding localization algorithm based on the BS information. Besides, this work maximized the estimation accuracy by assuming the fault of each element follows a specific distribution model.  However, in practice, the faults of the RIS elements may not follow a specific fault distribution. Furthermore, knowledge of a specific distribution is difficult to acquire  by the network manager, and its estimation may require a  significant number of  pilot signals.

Besides, it is worth noting that this challenge is difficult to address using the traditional optimization methods. Deep learning (DL)-aided  algorithms have emerged as a promising approach \cite{transfer_learning, Zhixiong1, Zhixiong2,densenet, Resnet}. DL algorithms are widely applied in wireless communication positioning due to their ability to model complex signal propagation in challenging environments \cite{3DCNN}, and their capacity for strong nonlinear mapping, which translates observed signal characteristics into precise location coordinates  \cite{CNNsurvey}.  However, the considered challenge remains non-trivial even when resorting to the powerful DL tools. {First, since the RIS is commonly comprised of a large number of elements, the detection of faulty elements across the RIS is   very challenging with existing DL algorithms, leading to the research gap of employing DL tools for faulty RIS elements detection.} Second, a specific neural network (NN) has to be designed for   reconstructing a complete high-dimensional signal. Third, the existing DL-aided localization algorithms require extensive training epochs and large datasets to ensure accurate estimation, making their deployment in dynamic environments unfeasible.

To address this challenge, we propose an effective localization algorithm based on the RIS information in the presence of   faulty elements.  {To address the detrimental effects of faulty elements, we are the first to harness transfer learning in devising an algorithm specifically aimed at detecting faulty elements within the RIS.} Then, to reconstruct the complete high-dimensional RIS information, we initially recover the  signal received by the non-faulty elements and subsequently restore the lost information from these identified faulty elements. Finally, to effectively localize the mobile user (MU) with smaller dataset and less training steps compared to the  traditional training-from-scratch DL algorithm, we employ transfer learning to design a localization algorithm. Our main contributions are summarized as follows:
\begin{itemize}
		\item[1)] {\textbf{\emph{High-accuracy faulty element detection}}: Given an RIS possessing some unknown  faulty elements, we harness transfer learning to design a two-phase transfer learning (TPTL) algorithm, addressing the faulty element detection problem. To cope with the limitations of multi-label classification, in \emph{Phase I}, the well-trained DenseNet 121 is leveraged to pinpoint faulty sub-arrays (SAs).  Building on this, in \emph{Phase II}, we  detect the faulty reflecting elements contained within the previously identified faulty SAs by transferring the NN of \emph{Phase I}. }

		\item[2)] {\emph{\textbf{Signal reconstruction on faulty RIS}}: We reconstruct the high-dimensional signal received at the RIS which contains rich information and can be utilized for high-accuracy localization. Specifically, based on the low-dimensional BS signal, we employ both a CNN and a VAE to recover the signal received at the non-faulty RIS elements and to restore the lost information at the faulty RIS elements. }

\item[3)] { \emph{\textbf{Effective localization with reconstructed signal}}:  We propose a transfer enhanced dual-stage (TEDS) localization algorithm, where the signal reconstruction is performed in \emph{Stage I} and the reconstructed signals are stored as fingerprints. In \emph{Stage II}, we transfer DenseNet 121 to estimate the coordinates of the MU. By fully exploiting the  RIS's high-dimensional information,  the proposed TFDS algorithm effectively improves the localization performance.
}

\item[4)] {\emph{\textbf{Direct fingerprint-based algorithm}}: As a benchmark algorithm,  we design a  transfer-enhanced direct fingerprint  (TEDF)  localization algorithm, which only exploits the BS information (e.g., received signal at the BS). The comparison between this algorithm and the TEDS algorithm allows us to better understand the benefits of using RIS information for localization.  }

\item[5)] { \emph{\textbf{Confirm the benefit of exploiting RIS information}}: {Through accurate faulty element detection and the effective signal reconstruction, our simulation results demonstrate that the localization performance of the proposed algorithm can benefit from the high-dimensional RIS information even with faulty elements. Additionally, the proposed TEDS localization algorithm showcases remarkable robustness against unoptimized phase shifts and signal-to-noise ratio (SNR) variations, with a marginal gap of only 0.001 between   `Opt. RIS, \(N_{\textrm{fau}}=0\)' and `RIS Inf., \(N_{\textrm{fau}}=0\)', emphasizing the minimal impact of phase shift optimization on localization accuracy.}
}		
	\end{itemize}
\section{System Model and Problem Formulation} \label{System_Model}
Consider an  RIS-aided  uplink (UL) localization system, where an  MU transmits pilot signals to the BS. The objective is to determine the MU's location with the aid of an RIS with some faulty elements but it is unknown which elements are damaged.  Furthermore, the BS is equipped with a uniform planar array (UPA) having $M_1\times M_2=M$ antennas, and the MU is equipped with a single antenna.  Moreover, the RIS is equipped with a UPA having $N_1\times N_2=N$ reflecting elements.
\subsection{Geometry and Channel Model}
The BS  is   located at coordinates $\bm{p}_b = [x_b, y_b, z_b]^T$, while the center of the RIS is situated at $\bm{p}_r = [x_r, y_r, z_r]^T$. The true position of the MU is denoted as $\bm{p}_u = [x_u, y_u, z_u]^T$. Typically, once  RIS and BS are deployed, their positions $\bm{p}_r$ and $\bm{p}_b$, are known and remain fixed. To motivate the deployment of an RIS,  we consider a classical scenario where the line-of-sight (LoS) path between the BS and the MU is blocked \cite{Zhi}. Such blockages can arise from numerous  factors, including buildings, vehicles, or natural obstructions \cite{Wuqq1}.

 Next, we model the MU-RIS and RIS-BS links. For any antenna array, given an elevation angle $\theta \in (0, \pi]$ and an azimuth angle $\phi \in (0, \pi]$, the array response vector can be generally described as
\begin{align}\label{gen_formula}
{\bm a}_{x}(\theta,\phi)={\bm a}_{x}^{(e)}(\theta)\otimes{\bm a}_{x}^{(a)}(\theta,\phi),
\end{align}
where $x$ represents the specific link and $\otimes$ denotes the Kronecker product, and
\begin{align}
{\bm a}_{x}^{(e)}(\theta) &= \left[1, \ldots, e^{\frac{-j2\pi(N_{e}-1) d_x\cos\theta}{\lambda_c}}\right]^T, \\
{\bm a}_{x}^{(a)}(\theta,\phi) &= \left[1, \ldots, e^{\frac{-j2\pi(N_{a}-1) d_x\sin\theta\cos\phi}{\lambda_c}}\right]^T,
\end{align}
where $N_{e}$ and $N_{a}$ denote the number of elements in the elevation and azimuth planes, respectively, $d_x$ denotes the distance between adjacent array elements, and $\lambda_c$ is the carrier wavelength.

\subsubsection{MU-RIS link}

Considering $P$ propagation paths between the MU and the RIS, and utilizing the general array response vector defined in   \eqref{gen_formula}, the channel response of the MU-RIS link, ${\bm g}_{ur}$, can be expressed as
\begin{align}
{\bm g}_{ur} = \sum^{P}_{p=1}\alpha_{p}{\bm a}_{R_a}(\theta_{p},\phi_{p}),
\end{align}
where ${\bm a}_{R_a}$ denotes the array response vector of the RIS and $\alpha_{p}$ represents the channel gain of the $p$-th path.

\subsubsection{RIS-BS link}

Assuming $J$ propagation paths between the RIS and the BS, the channel matrix of the RIS-BS link, ${\bm H}_{rb}$, can be modeled as
\begin{align}
{\bm H}_{rb}=\sum_{j=1}^{J}\beta_{j}{\bm a}_{B}(\theta_{j},\phi_{j}){\bm a}^{H}_{R_d}(\psi_{j},\omega_{j}),
\end{align}
where $\beta_{j}$ represents the channel gain of the $j$-th path and ${\bm a}_{B}$ is the array response vectors of the BS.

\subsection{Faulty Element Model}
The RIS panel is  passive, lacking the self-diagnostic capabilities of active antennas. Given this absence of direct feedback, establishing an external detection mechanism becomes crucial. For this purpose, drawing from the faulty antenna model in \cite{Babur}, we introduce a corresponding model for the RIS elements as follows.
\subsubsection{RIS Faulty Element Model}
Without faulty elements, the RIS phase shift vector is defined as ${\bm \omega} = [\omega_1, \cdots, \omega_N]^T \in \mathbb{C}^{N\times1}$. Actually, some of the reflecting elements may be damaged (refer to as faulty elements in this paper) due to some unknown environment conditions, leading to the situation that they cannot effectively receive/reflect any signals \footnote{There may be other metrics to define whether an RIS element is damaged or not, which will be left to our future work.}.

 We assume that the received signal strength at the $n$-th element of the RIS is denoted as $y^{(r)}_n$. Besides, we employ  a threshold parameter $\zeta_r\ll p_r$ to determinate the fault status of the $n$-th element, where $p_r$ represents the minimum transmit power to the reflecting elements, and $\zeta_r$ is assumed to be exceedingly  small. Accordingly, the fault status is mathematically defined as
\begin{align}\label{13}
\mathcal{B}_n =\begin{cases}
	0, y^{(r)}_n\leqslant \zeta_r\\
	1, y^{(r)}_n>\zeta_r, \quad n\in\{1,\cdots,N\}.\\
\end{cases}
\end{align}
As we can see from \eqref{13}, there are two distinct cases: First,  when $y^{(r)}_n \leqslant \zeta_n$, this indicates that the strength of the  RIS received signal is below or equal to the threshold $\zeta_n$, suggesting that this element is damaged. Consequently, we assign $\mathcal{B}_n = 0$ in such cases.  Conversely, when  the received signal at the RIS exceeds the threshold $\zeta_n$, i.e., $y^{(r)}_n > \zeta_n$, this  signifies that the $n$-th element is operating correctly. Consequently, we assign $\mathcal{B}_n = 1$ in these situations.

Let us integrate the statuses of all elements into a vector as follows
   \begin{align}\label{14}
\bm{\mathcal{B}}= [\mathcal{B}_1,\cdots,\mathcal{B}_N]^T.
\end{align}
Accordingly,  the RIS  phase shifts profile can be mathematically formulated  as follows
   \begin{align}\label{15}
{\bm \varpi}={\bm \omega}\odot \bm{\mathcal{B}}
 = [\omega_1\mathcal{B}_1,\cdots, \omega_N\mathcal{B}_N]^T
=[\varpi_1,\cdots,\varpi_N]^T,
\end{align}
where $\odot$ denotes the Hadamard product.

\subsection{Signal Model}
{The UL signal received by the RIS can be expressed as
\begin{align}\label{aa16}
{\bm y}_r= {\bm g}_{ur}{ s}.
\end{align}}
Accordingly, The UL signal received by the BS can be expressed as
\begin{align}\label{16}
{\bm y}={\bm H}_{rb}\textrm{diag}({\bm \varpi}){\bm g}_{ur}{ s}+{\bm n},
\end{align}
where $\textrm{diag}({\bm \varpi})$, ${s}$, and ${\bm n}$ denote the phase shift matrix of the RIS, the pilot signal transmitted by the MU, and the zero-mean additive white Gaussian noise, respectively.
\section{Problem Formulation}
Building upon the received signal ${\bm y}$, defined in \eqref{16}, our objective is to design a localization algorithm for the scenario where the RIS contains deterministic faulty elements. To this end, it is essential to   detect the faulty elements, which enables us to mitigate their harmful impact on the localization accuracy.
 Unlike active components that can self-diagnose or report inconsistencies, the potential faults or damages within the RIS are invisible without external intervention. Detecting and localizing these faulty elements requires a deeper understanding and innovative methodologies, presenting a non-trivial challenge in the field.   In the following subsection, we will delve formulate a faulty elements detection problem.
\subsection{Faulty Element Detection Problem}
First, let \(\bm{\hat{\mathcal{B}}}\) denote an estimate of \(\bm{\mathcal{B}}\), given by
\begin{align}\label{17}
\bm{\hat{\mathcal{B}}} = \mathcal{F}({\bm y}) = [\hat{\mathcal{B}}_1, \ldots, \hat{\mathcal{B}}_N]^T,
\end{align}
where \(\mathcal{F}(\cdot)\) denotes the non-linear function between \({\bm y}\)  and \(\bm{\hat{\mathcal{B}}}\), and \(\hat{\mathcal{B}}_n\)  represents the estimated status of the \(n\)-th RIS element, which is either \(1\) (functioning) or \(0\) (faulty). Consequently, the detection problem is given by
\begin{align}\label{18}
\textrm{({P}1)}: \underset{\bm{\hat{\mathcal{B}}}}{\min} \ \mathcal{L}_{\mathcal{B}} \left( \mathcal{F} \left( \boldsymbol{y} \right) , \boldsymbol{\mathcal{B}} \right),
\end{align}
where \(\mathcal{L}_{\mathcal{B}}\) is the loss function for predicting the element status.   To further illustrate the proposed problem and the related challenge, we make the following remarks.
  \begin{figure*}
    \centering
    \includegraphics[width=0.75\linewidth]{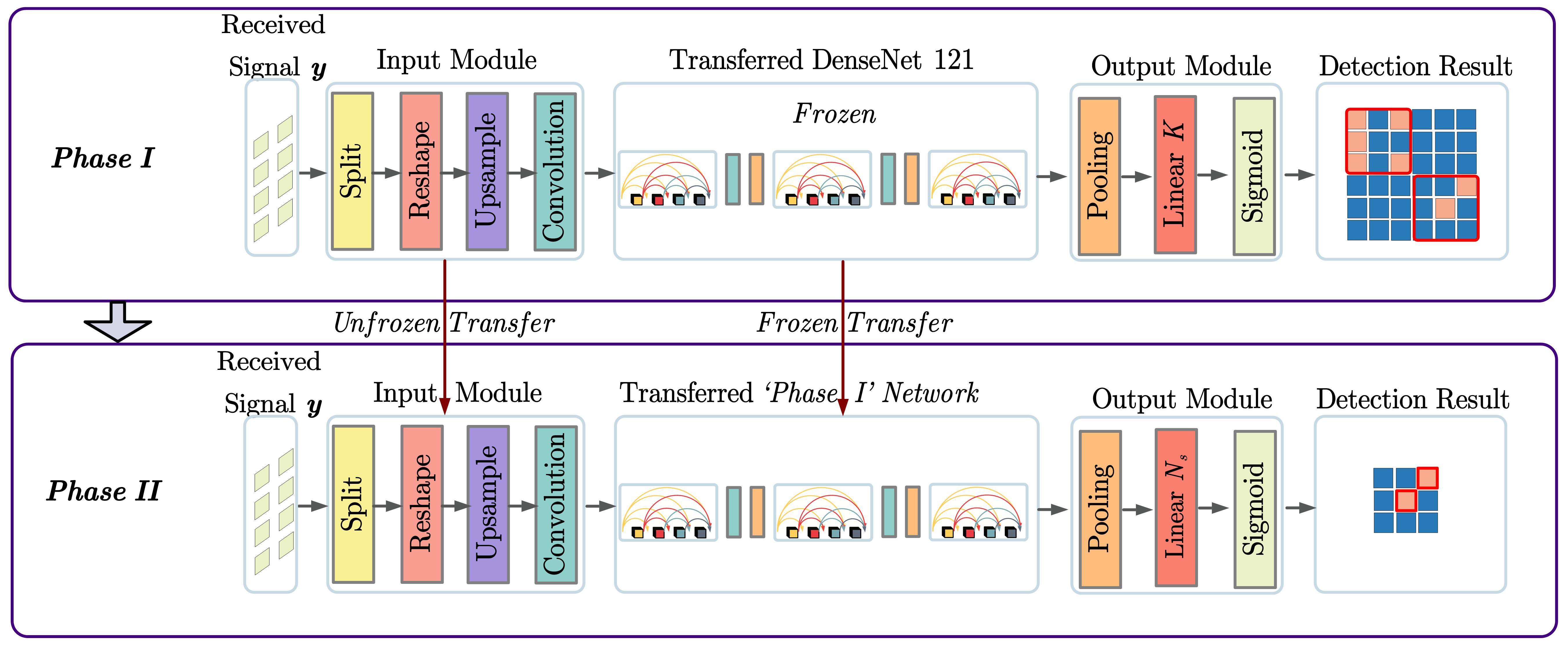}
    \caption{Network structure of  TPTL algorithm.}
    \label{TPTL}
\end{figure*}
\begin{remark} \label{rm1}
Given the nature of the received signal $\bm{y}$ at the BS, $\bm{y}$ can be regarded as an image, drawing parallels between problem (P1) and tasks in computer science (CS) \cite{CNNsurvey}. Since the status of the  individual elements, ${\mathcal{B}}_{n}$, is either $1$ or $0$, the estimation of the fault status corresponds to a binary classification problem. Furthermore, considering that the RIS comprises multiple elements, (P1) can be regarded as a multi-label classification problem.
\end{remark}

\begin{remark}
Contemporary research on multi-label classification has only advanced to handle datasets with up to $100$ labels \cite{liu2021query2label}. For instance, datasets like ``MS-COCO" encompass $122,218$ images spanning $80$ labels \cite{coco2023}. However, the number of the reflecting elements of an RIS can easily surpass $100$  (e.g., $N_1\times N_2> 10\times 10$). In other words, the number of the labels that need to be classified will exceed $100$ and reach thousands, posing a significant challenge for detecting faulty elements using the existing DL algorithms.
\end{remark}
To address the challenge in \textbf{Remark 2}, we propose to divide the RIS panel into multiple SAs, each containing a same number of reflecting elements. Then, Problem (P1) is decomposed into two sub-problems. The first sub-problem is to detect the SA containing the faulty elements (referred to as the  ``faulty SA"), represented by Problem (P1-a). The second sub-problem is to detect the faulty elements within the faulty SA, represented by Problem (P1-b). These two sub-problems will be sequentially introduced in the following.
\subsubsection{Detection of Faulty SA}
To begin with, let us introduce the procedures for dividing and labeling the SAs. The RIS with $N$ elements is divided  into $K$ SAs, with each SA containing $N_s=\lfloor N/K \rfloor$  reflecting elements \footnote{$\lfloor x \rfloor$ denotes the function that output the greatest integer less than of equal to $x$.}. If the $k$-th SA contains one or more faulty reflecting elements,  we classify it as a faulty SA. Conversely, if the $k$-th SA does not contain any faulty   elements, we determine the $k$-th SA to be operating correctly. Then, by denoting the number of faulty   elements of the $k$-th SA as $N_{k}^{(\textrm{faulty})}$,  the status of the $k$-th SA can be mathematically defined as
\begin{align}\label{19}
\mathcal{C}_k =\begin{cases}
	0, N_{k}^{(\textrm{faulty})}\geq 1,\\
	1, N_{k}^{(\textrm{faulty})}=0, \quad k\in\{1,\cdots,K\}.\\
\end{cases}
\end{align}
By collecting the status of all SAs in vector form, we have $\bm{\mathcal{C}}= [\mathcal{C}_1,\cdots,\mathcal{C}_K]^T$. The estimate of $\bm{\mathcal{C}}$ is thus defined as $\hat{\bm{\mathcal{C}}}=\mathcal{X}({\bm y})= [\hat{\mathcal{C}}_1,\cdots,\hat{\mathcal{C}}_K]^T$,
where $\mathcal{X}(\cdot)$ denotes a non-linear function and $\hat{\mathcal{C}}_k$ denotes an estimate of ${\mathcal{C}}_k$.

Building upon this, the problem of detecting the faulty SAs can be formulated as follows
\begin{align}\label{22}
\textrm{(P1-a)} \quad \underset{\hat{\bm{\mathcal{C}}}}{\min}\quad\mathcal{L}_{\mathcal{C}} \left( \mathcal{X} \left( \boldsymbol{y} \right) ,\boldsymbol{{\bm{\mathcal{C}}} } \right),
\end{align}
where $\mathcal{L}_{\mathcal{C}}$ denotes the specific loss function for estimating the status of all SAs.
\subsubsection{Detection of Faulty Elements of SA}
After identifying  the faulty SAs, the next step is to detect the faulty elements within a given faulty SA. Similar to \eqref{14}, the statuses of the faulty elements within the $k$-th faulty SA can be organized in a vector,  denoted as $\bm{\mathcal{B}}^{(k)}_{sub}= [\mathcal{B}_1,\cdots,\mathcal{B}_{N_s}]^T$.
Thus, the estimate of $\bm{\mathcal{B}}^{(k)}_{sub}$  can be written as $\hat{\bm{\mathcal{B}}}^{(k)}_{sub}= [\hat{\mathcal{B}}^{(k)}_1,\cdots,\hat{\mathcal{B}}^{(k)}_{N_s}]^T$, where $\hat{\mathcal{B}}^{(k)}_{n_s}$, $ n_s\in\{1,\cdots,N_s\}$, represents the estimated status of the $n_s$-th element within the $k$-th faulty SA. Besides, we have $\mathcal{H}(\cdot)$ to denote  the complex non-linear function between the received signal $ {\bm y}$ and the estimated statuses of faulty elements within the $k$-th faulty SA.

Consequently, detecting the faulty elements within the $k$-th faulty SA  can be formulated  as follows
\begin{align}\label{25}
\textrm{(P1-b)} \quad \underset{\hat{\bm{\mathcal{B}}}^{(k)}_{sub}}{\min}\quad\mathcal{L}_{\mathcal{B}_{sub}} \left( \mathcal{H} \left( \boldsymbol{y} \right) ,\boldsymbol{\mathcal{B} }^{(k)}_{sub} \right),
\end{align}
where $\mathcal{L}_{\mathcal{B}_{sub}}$ denotes the specific loss function for predicting the statuses of the faulty elements within  the $k$-th faulty SA.

 Similar to ({P}1), these two sub-problems can also be regarded as a kind of multi-label classification problem. In the field of CS, the multi-label classification problem can be solved by using a CNN \cite{CNNsurvey}. Notably, the parameters of the CNN have to be trained with a large number of data sets before being used in the online phase \cite{CNNsurvey}. However, training two CNN networks to solve the above problems is inefficient. To address this challenge and to save time and computation resources, we propose to leverage transfer learning algorithms \cite{transfer_learning} to fine-tune  pre-trained CNN networks.

\subsection{Localization Problem Under Faulty Elements}
The localization accuracy of  existing localization algorithms designed under assumption of a perfect RIS is expected to deteriorate in the presence of faulty RIS elements. Therefore, building upon the obtained knowledge on the faulty elements, i.e., $\hat{\bm{\mathcal{ B}}}$, it is crucial to develop a localization algorithm rather that can correct the harmful impact of faulty elements.  By  defining the estimate of the position of the MU as $\bm{\hat{p}}_u$,   the relationship with the knowledge of faulty elements can be expressed as
\begin{align}\label{26}
\bm{\hat{p}}_u=\mathcal{Q}({\bm y},\bm{\hat{\mathcal{B}}})= [\hat{x}_u, \hat{y}_u, \hat{z}_u]^T,
\end{align}
where $\mathcal{Q}(\cdot)$ represents the intricate non-linear mapping between received signal $ {\bm y}$ and the estimated position of the MU, $\bm{\hat{p}}_u$.
In the next step, we tackle the localization problem, labeled as ({P}2). This problem aims to accurately predict the 3D position of the MU by minimizing the loss function, $\mathcal{L}_{\bm{p}}$. Mathematically, it is formulated as
\begin{align}\label{27}
\textrm{({P}2)}  \quad \underset{\bm{\hat{p}}_u }{\min} \ &\mathcal{L}_{\bm{p}} \left( \mathcal{Q}({\bm y},\bm{\hat{\mathcal{B}}}),\bm{{p}}_u \right).
\end{align}
Traditional convex optimization and channel estimation methods, as highlighted in \cite{WuFingerprint}, are not capable of accurately estimating $\bm{{p}}_u$. Therefore, this paper introduces a fingerprint-based algorithm for this purpose. After   the faulty elements have been identified, we reconstruct from the low-dimensional BS signal to complete high-dimensional RIS signal. This reconstructed RIS received signal then acts  as a unique fingerprint for estimating $\bm{{p}}_u$. For comparison, we also utilize the BS received signal as another fingerprint to estimate $\bm{{p}}_u$.
\section{Transfer Learning Empowered Faulty Element Detection Algorithm Design}
\begin{figure*}[t!]
   \centering
		\includegraphics[width=1.0 \linewidth]{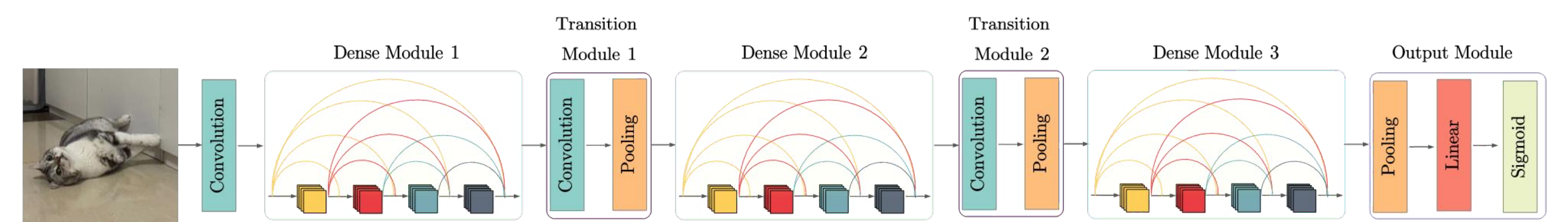}
		\caption{\normalsize Network structure of  DenseNet 121.}\label{DenseNet_121_Model}
\end{figure*}
In this section, we aim to solve Problem ({P}1) to estimate the fault statuses  of the RIS elements which serves as the foundation for the design of the proposed localization algorithms. To this end,  we propose a TPTL algorithm, whose the network structure is depicted in Fig. \ref{TPTL}. TPTL employs transfer learning in two distinct phases to sequentially tackle problems (P1-a) and (P1-b). Before introducing the proposed TPTL, we first explain why the transfer learning \cite{transfer_learning} is adopted in the following subsection.
\subsection{Adoption  of Transfer Learning}
Transfer learning \cite{transfer_learning} is particularly advantageous for tasks like detecting faulty elements within the  RIS. One of the primary reasons behind its suitability is the analogous nature of received signals to images. Received signal ${\bm y}$ can be viewed as an image, also with features to be extracted and recognized. This similarity allows us to harness the power of models pre-trained for image recognition tasks and adapt them for   faulty elements detection.  Transfer learning can efficiently use data from image domains to facilitate  faulty elements detection.

Hence, we propose a TPTL algorithm to sequentially address problems (P1-a) and (P1-b). As shown in Fig. \ref{TPTL}, during  \emph{Phase I}, we utilize the classical DenseNet 121 \cite{densenet} as the base model for transfer learning, adapting the input and output modules for Problem (P1-a).  Upon completing \emph{Phase I} training, we save the  trained NN to serve as the base model for  transfer learning in \emph{Phase II}.   Furthermore, in \emph{Phase II}, we make necessary adjustments to both the input and output modules of the well-trained model to better fit Problem (P1-b). To facilitate an intuitive understanding of the employed approach, the details of DenseNet 121 will be introduced in the following subsection.
\subsection{The Architecture of DenseNet 121}
DenseNet \cite{densenet}, inspired by the foundational principles of ResNet \cite{Resnet}, reuses the features from earlier layers. However, different from the skip connections  which might bypass a couple of layers, DenseNet establishes connections between every layer and every other layer, ensuring a more robust feature propagation. This approach, while reducing the parameter count, allows DenseNet to potentially surpass the performance of ResNet. The design of DenseNet 121 is illustrated in Fig. \ref{DenseNet_121_Model}, and its architecture comprises three main components: Dense Modules, Transition Modules, and an Output Module, which will be introduced as follows.
\subsubsection{Dense Module}
\begin{figure}[t]
	\centering
	\includegraphics[width=0.3\textwidth]{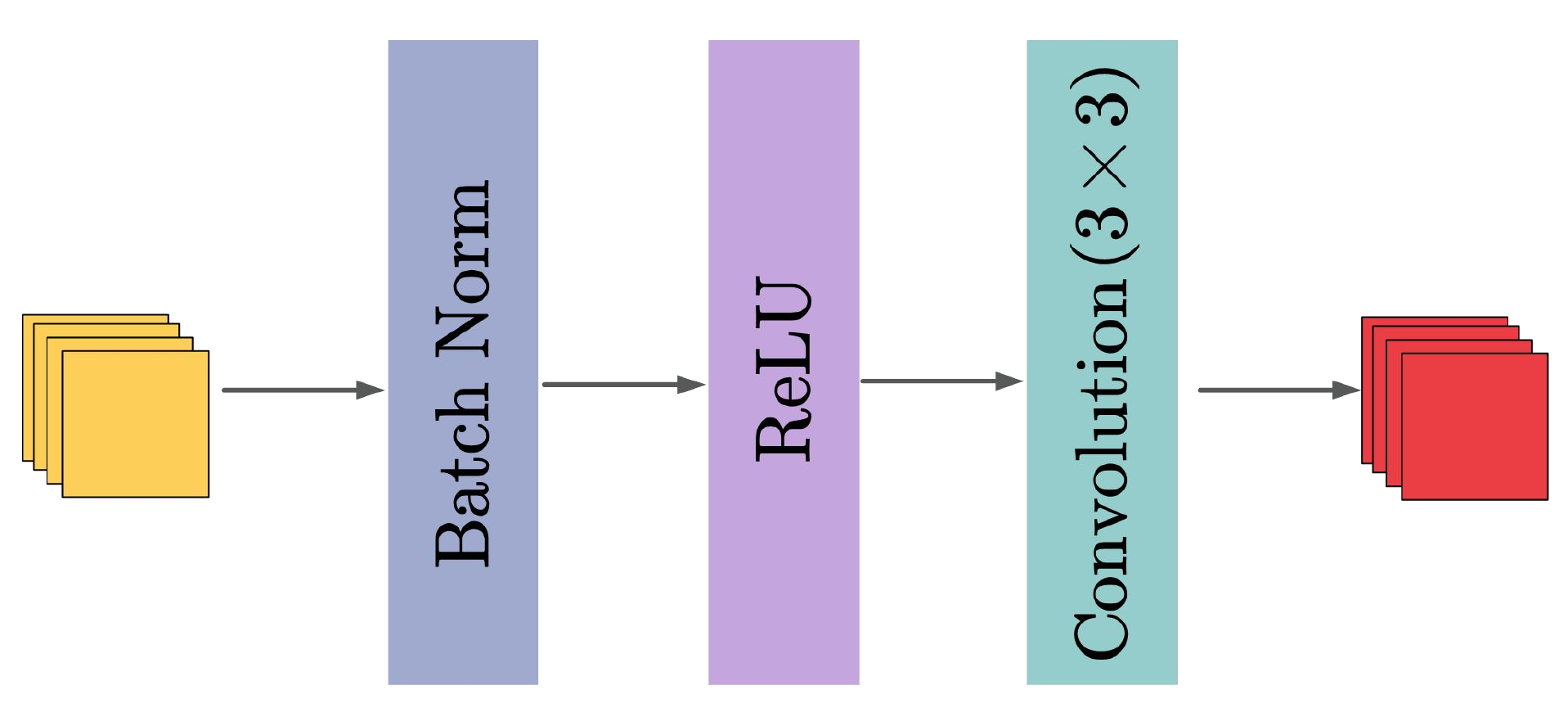}
	\caption{Network structure of a Dense Block.}
	\label{Dense_Module}
\end{figure}

As depicted in  Fig.  \ref{DenseNet_121_Model}, the Dense Module comprises multiple Dense Blocks. The details of a Dense Block are shown in Fig.  \ref{Dense_Module}. Each Dense Block starts with a batch normalization (BN) layer, followed by a layer rectified linear unit (ReLU) activation function layer, and concludes with a $3\times3$ convolutional (Con) layer. A notable characteristic of the Dense Module is its connectivity: Every block incorporates the feature maps of all preceding blocks within the module.  This design ensures efficient feature reuse and a rich flow of information throughout the network \cite{densenet}.

\subsubsection{Transition Module}
From Fig. \ref{DenseNet_121_Model}, we observe that the \emph{Transition Module}, which acts as a connector between \emph{Dense Modules}, starts with a $1\times1$ Con layer, followed by a $2\times2$ average pooling layer, effectively reducing the spatial dimensions of the feature maps and streamlining them for either the next \emph{Dense Module} or the concluding output layer.
\subsubsection{Output Module}
 This module employs average pooling to aggregate the feature maps. Subsequently, these integrated features are channeled through a linear layer. A sigmoid activation function then finalizes the process, ensuring that the output values lie between $0$ and $ 1$.

Having detailed the architecture of DenseNet 121, we will now explain the proposed TPTL algorithm, which is structured in two distinct phases. These phases are elaborated in the following subsections.
\subsection{Phase I: Transfer Learning of DenseNet 121}
To extract the features from the received signal $\boldsymbol{y}$ and tackle sub-problem (P1-a), we employ transfer learning \cite{transfer_learning} with a pre-trained DenseNet 121 \cite{densenet}. Though the internal layers of DenseNet 121 are effective at feature extraction due to their extensive pre-training on datasets such as ImageNet \cite{deng2009imagenet}, adjustments are necessary to cater to the specific nuances of sub-problem (P1-a) and the nature of signal $\boldsymbol{y}$. As shown in Fig. \ref{TPTL}, an \emph{Input Module} is introduced to preprocess the received signal, while the \emph{Output Module} is configured to classify the different statuses of the SAs on the RIS panel. Additionally, choosing an appropriate loss function for this binary classification problem is crucial.   The details will be discussed in the subsequent paragraphs.
\subsubsection{Input Module}\label{input_module}
The received signal at the BS, i.e., $\boldsymbol{y}$,  is first processed by the Input Module so that it can be fed into the transferred DenseNet 121 for classifying the statuses of the SAs \cite{transfer_learning}.  To this end, we first decompose signal $\boldsymbol{y}$  as follows
\begin{equation}
\boldsymbol{y} = \mathfrak{R}(\boldsymbol{y}) + j\mathfrak{I}(\boldsymbol{y}),
\end{equation}
where $\mathfrak{R}(\boldsymbol{y})$ and $\mathfrak{I}(\boldsymbol{y})$ are  the real and imaginary parts, respectively. In this way, we can ensure that both the real and imaginary information contributes to enhancing signal differentiation and facilitating feature extraction.

Thereafter, we reshape both $\mathfrak{R}(\boldsymbol{y})$ and $\mathfrak{I}(\boldsymbol{y})$ into matrices, which are denoted as ${\bm M}_{Ry}\in \mathbb{R}^{ M_1 \times M_2}$ and ${\bm M}_{Iy}\in \mathbb{R}^{ M_1 \times M_2}$, respectively. The primary motivation for this particular reshaping strategy is to effectively extract meaningful features from the UPA of the BS, which spans both the horizontal and vertical dimensions.  Then, we concatenate the reshaped matrices of the real and imaginary parts, resulting in a 3D tensor represented as
\begin{equation}\label{ty}
\boldsymbol{T}_{y} = \begin{bmatrix}
{\bm M}_{Ry}  \\
{\bm M}_{Iy}
\end{bmatrix}\in \mathbb{R}^{2 \times M_1 \times M_2}.
\end{equation}
However, the values of $M_1$ and $M_2$ may not be directly compatible with DenseNet 121. Hence, an upsampling technique is applied:
\begin{equation} \label{tup}
\boldsymbol{T}_{\text{up}} = \text{Upsample}(\boldsymbol{T}_{y}) \in \mathbb{R}^{2 \times 256 \times 256}.
\end{equation}
The upsampling technique is employed to adjust the dimensions of the tensor $\boldsymbol{T}_{y} $  to the desired dimensions of $2 \times 256 \times 256$. It typically involves pixel value interpolation to expand the image or tensor size, redistributing pixel values accordingly. To make $\boldsymbol{T}_{\text{up}}$ suitable for processing by DenseNet 121 \cite{densenet}, which expects an input tensor of the form $(3,256,256)$, a Con Layer is introduced to transform the bi-channel tensor to a tri-channel one:
\begin{equation}\label{tdense}
\boldsymbol{T}_{\text{DenseNet}} = \text{ConLayer}(\boldsymbol{T}_{\text{up}}).
\end{equation}
\subsubsection{Transferred DenseNet 121}
$\boldsymbol{T}_{\text{DenseNet}}$ is input to the well-trained DenseNet 121 without the first Con layer and \emph{Output Module}. While the internal layers of DenseNet 121 are frozen during the early stages of training to maintain their feature extraction capabilities, they are gradually unfrozen later, enabling minor adjustments to adapt better to the specific task of detecting faulty SAs. This ensures that the network leverages the generic feature extraction capabilities of the pre-trained model while also fine-tuning specific layers to our dataset and problem definition for SA detection.
\subsubsection{Output Module}
In the typical use case, DenseNet 121 is employed for single-label image classification using a softmax layer to determine category probabilities \cite{softmax_vs_sigmoid}. However, our task in (P1-a) entails a multi-label binary classification problem concerning the statuses of the SAs. Specifically, we have $ K $ SAs, and for each one, the goal is to determine if it is faulty or not. The status of  a given SA is independent of the statuses of the other SAs. To accommodate this, we modify the linear layer of the \emph{Output Module} to produce $ K $ outputs, which is  aligned with the number of SAs under consideration.
Furthermore, we replace the softmax layer with the sigmoid layer \cite{softmax_vs_sigmoid}, which allows for individual probability estimation for each label. Defining the output from the linear layer by  $\boldsymbol{x}_\textrm{p1}$, the probability that the $k$-th SA is faulty (represented as $\hat{\mathcal{C}}_k = 0$) can be expressed as
\begin{align}
P(\hat{\mathcal{C}}_k = 0 | \boldsymbol{x}_\textrm{p1}) ={F}_{\textrm{sig}}(\boldsymbol{w}_k^{T} \cdot \boldsymbol{x}_\textrm{p1} + b_k),
\end{align}
where ${F}_{\textrm{sig}}$ denotes  the employed sigmoid function, $\boldsymbol{w}_k$ is the weight vector associated with the $k$-th SA, and $b_k$ represents the corresponding bias.
\subsubsection{Loss Function}
For addressing the complexities of our multi-label classification task, it is imperative to employ a suitable loss function. The \emph{multi-label soft margin loss} in \cite{neural_networks_multilabel} is employed as the specific loss function for addressing the multi-label classification problem in this paper. The application of the \emph{multi-label soft margin loss} is motivated by its design, specifically tailored for multi-label scenarios \cite{softmax_vs_sigmoid}. Unlike conventional loss functions, the \emph{multi-label soft margin loss} was constructed for problems where input samples might simultaneously belong to multiple labels, providing the means to individually estimate label probabilities. The mathematical formulation of the loss function is given by
\begin{equation}
L_{msml} = -\frac{1}{K} \sum_{k=1}^{K} [\mathcal{C}_k  \log(\hat{\mathcal{C}}_k) + (1 - \mathcal{C}_k) \log(1 - \hat{\mathcal{C}}_k)],
\end{equation}
where $K$ represents the total number of SAs (or labels) in the RIS. For the $k$-th SA, $\mathcal{C}_k$ denotes the true label, while $\hat{\mathcal{C}}_k$ represents the predicted label.

\subsubsection{Training for Faulty SA Detection}
Once the \emph{Input Module}, \emph{Output Module}, and loss function adjustments are established for the SA classification, the next step is to fine-tune the pre-trained DenseNet 121 on our specific dataset for detecting faulty SAs.  This means that we train the model on a new dataset specifically focused on detecting faulty SAs. The utilization of weights initialized from prior training on comprehensive datasets, such as ImageNet \cite{deng2009imagenet}, offers a foundation that aids faster convergence and potentially better generalization \cite{transfer_learning}.

The primary objective during the training step is to optimize the network weights by minimizing the loss $L_{\mathcal{C}}$, which is specifically designed for the faulty SA detection problem (P1-a). By training and converging the network, Problem (P1-a)  of estimating the faulty SAs of the RIS can be solved. The training process involves iteratively adjusting the network weights based on the gradient of the loss with respect to the weights. Mathematically, the optimization problem for training in \emph{Phase I}  can be expressed as
\begin{equation}
\min_{\boldsymbol{\theta}_\textrm{P1-a}} L_{\mathcal{C}}(\boldsymbol{\theta}_\textrm{P1-a}),
\end{equation}
where $\boldsymbol{\theta}_\textrm{P1-a}$ represents the parameters (or weights) of the designed network for the detection of faulty SAs in \emph{Phase I}.

\subsubsection{Output after Training}
Upon training convergence, the network outputs  undergo the sigmoid activation function as detailed previously. This produces probabilities for each status of the faulty SA. The classification based on a threshold of $0.5$ is given by
\begin{equation}
\hat{\mathcal{C}}_k = \begin{cases}
0 & \text{if } P(\hat{\mathcal{C}}_k = 0 | \boldsymbol{x}_\textrm{p1}) \geq 0.5; \\
1 & \text{otherwise}.
\end{cases}
\end{equation}
Besides, the threshold can be tailored to meet precision-recall needs or practical requirements \footnote{  {
 To adjust the classification threshold, we can analyze precision-recall curves to find a balance between precision and recall suitable for our application. A higher threshold is used to prioritize precision, while a lower threshold is chosen to favor recall.  }} .

\subsection{Phase II: Transfer Learning of `Phase I' Network}
 { Problem (P1-b)  is similar to Problem (P1-a), since they have the same prior knowledge $\boldsymbol{y}$ and are both muti-label classification tasks. }  Hence,  transfer learning is employed in this phase to reduce the training time and cost. Specifically, we adopt the well-trained NN  from \emph{Phase I} as the base model for transfer learning in \emph{Phase II}.

 If $\mathcal{F}_1$ denotes the feature space shaped in \emph{Phase I} and $\mathcal{F}_2$ represents the desired feature space for \emph{Phase II}, our intent is to discern a transformation function $ \phi: \mathcal{F}_1 \rightarrow \mathcal{F}_2 $ that minimizes the divergence between these spaces, which is expressed as
\begin{equation}
\label{eq:transfer}
\phi^* = \arg\min_{\phi} \lVert \mathcal{F}_1 - \phi(\mathcal{F}_2) \rVert^2.
\end{equation}

Leveraging the NN  pre-trained in \emph{Phase I} as the base model, we introduce modifications tailored for \emph{Phase II}, especially in the \emph{Output Module}. By doing so,  we can ensure that the foundational features and representations learned in \emph{Phase I} remain intact while the network is trained to better align with the task of \emph{Phase II}. The network structure of \emph{Phase II} is shown in Fig. \ref{TPTL}. Details of the modifications are presented in the following.
\subsubsection{Input Module of Phase II}
Although \emph{Phase II} directly transfers the input module from \emph{Phase I}, it is essential to \emph{unfreeze} this module before training. This is due to the distinct tasks associated with each phase. This operation aids the network to adapt more effectively to the new problem.

\subsubsection{Output Module}
To tackle Problem (P1-b), it is imperative to make adjustments to the linear layer within the \emph{Output Module}. Specifically, the output dimension of this linear layer needs to be adjusted to $N_s$. Hence, the \emph{Output Module} can yield the  probabilities associated with the fault  statuses of the $N_s$ reflecting elements in the $k$-th SA. Denoting  the output of this modified linear layer by $\boldsymbol{x}_\textrm{p2}$, the probability of the $n_s$-th reflecting element in the $k$-th SA being faulty (denoted as $\hat{\mathcal{B}}_{n_s}^{(k)} = 0$) can be expressed as
\begin{align}
P(\hat{\mathcal{B}}_{n_s}^{(k)}  = 0 | \boldsymbol{x}_\textrm{p2}) ={F}_{\textrm{sig}}(\boldsymbol{w}_{n_s}^{T} \cdot \boldsymbol{x}_\textrm{p2} + b_{n_s}),
\end{align}
where  $\boldsymbol{w}_{n_s}$ denotes the weight vector associated with the ${n_s}$-th reflecting element and $b_{n_s}$ represents the corresponding bias.

\subsubsection{Training for Phase II}
Once the adjustments have been made,  the next step is fine-tuning the pre-trained NN of \emph{Phase I} on another specific dataset for detecting faulty reflecting elements within each faulty SA. This means that we have to train the model on a new dataset for detecting faulty reflecting elements.

The primary objective during the training step is to optimize the network weights by minimizing the loss $\mathcal{L}_{\mathcal{B}_{sub}}$, which is specifically designed for the faulty reflecting elements detection Problem (P1-b). By training and converging the network using this optimization, Problem (P1-b) of estimating the faulty reflecting elements can be solved.  Mathematically, the optimization problem for training in \emph{Phase II}  can be formulated as
\begin{equation}
\min_{\boldsymbol{\theta}_\textrm{P1-b}} \mathcal{L}_{\mathcal{B}_{sub}}(\boldsymbol{\theta}_\textrm{P1-b}),
\end{equation}
where $\boldsymbol{\theta}_\textrm{P1-b}$ denotes the parameters (or weights) of the network for this specific sub-problem. Using the gradient of the loss with respect to these weights, the training process iteratively adjusts the network weights until convergence.

Crucially, for \emph{Phase II}, the internal layers of the NN transferred from \emph{Phase I} are initially frozen during the early stages of training to retain their feature extraction prowess. However, as training progresses, these layers are gradually unfrozen, allowing for   better adaptation to the specific task of detecting faulty reflecting elements. This allows the network to use the pre-trained model's general features and fine-tune specific layers for detecting faulty reflecting elements.

\subsubsection{Output after Training}
Upon training convergence, the network outputs  are fed into a the sigmoid activation function. This produces probabilities for each status of the reflecting elements within the SA. A threshold, typically set at $0.5$, determines the binary classification:
\begin{equation}
\hat{\mathcal{B}}_{n_s}^{(k)} = \begin{cases}
0 & \text{if } P(\hat{\mathcal{B}}_{n_s}^{(k)} = 0 | \boldsymbol{x}_\textrm{p2}) \geq 0.5, \\
1 & \text{otherwise}.
\end{cases}
\end{equation}
The threshold can be adjusted based on specific precision-recall requirements or based on the practical requirements.

\subsection{Trained TPTL Algorithm Usage Guide}
After  training of the TPTL algorithm, we integrate the well-trained models into a `sealed black box'. Below are the instructions for using the TPTL algorithm to detect the position of faulty elements on the RIS.
\subsubsection{Phase I (Diagnosis of Faulty SAs)}
 The MU acts as a diagnostic device, transmitting a pilot signal towards the BS via the RIS. The received signal at the BS is then input to the well-trained model of \emph{Phase I} to output the statuses of the SAs within the RIS panel.

\subsubsection{Phase II (Diagnosis of Faulty Elements within Detected SA)}
After identifying a faulty SA, all other SAs are turned off, and the MU sends another pilot signal to the BS. The received signal at the BS is then input to the well-trained model of \emph{Phase II} to estimate the faulty statuses of the elements within this faulty SA.  { This method effectively utilizes the comprehensive information provided by the  BS received signal  to distinguish between varying faulty scenarios across SAs, thereby demonstrating the versatility and robustness of our proposed diagnostic approach in \emph{Phase II}.}

By sequentially addressing Problem (P1-a) and  Problem (P1-b) with the TPTL algorithm, the overall Problem (P1) is resolved.
\section{Transfer-Enhanced Dual-Stage  Algorithm: RIS Signal Reconstruction and Localization}\label{TEDSA}
\begin{figure*}
    \centering
    \includegraphics[width=0.9\linewidth]{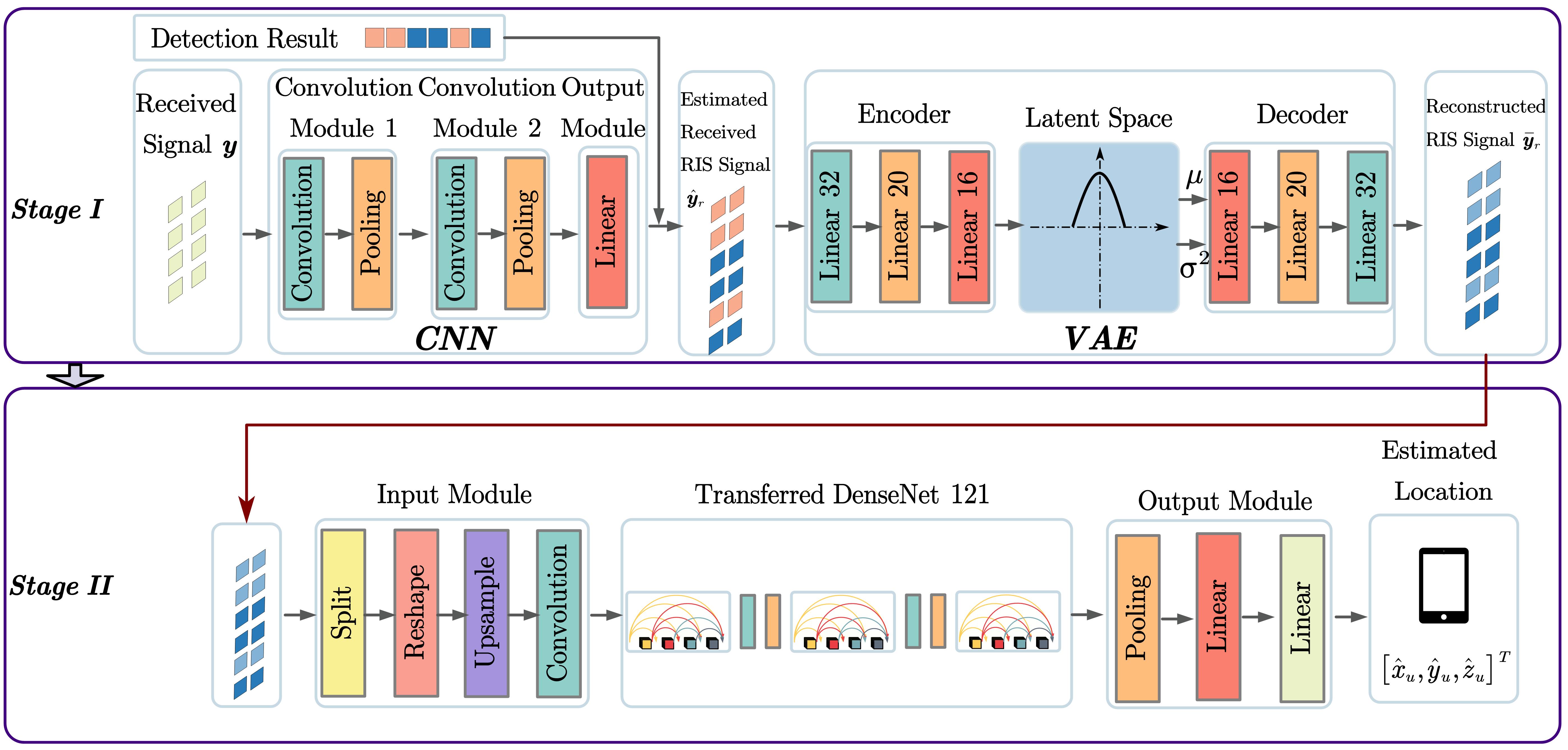}
    \caption{Network structure of   TEDS algorithm.}
    \label{TEDS}
\end{figure*}
After having identified the faulty elements, our primary aim is to mitigate their detrimental effects and fully realize the potential of high-dimensional RIS information for localization. Addressing this, we introduce the proposed  TEDS algorithm, whose network structure is depicted in Fig. \ref{TEDS}. Deriving a closed-form solution for the 3D coordinates of the MU using traditional optimization methods is generally impossible \cite{WuFingerprint}. Fingerprint-based methods emerge as an enticing solution, utilizing the received signal as a unique fingerprint to determine the MU's location \cite{Xiqi}. The proposed TEDS algorithm employs the fingerprint-based method, exploiting the high-dimensional RIS information which includes rich information beneficial for localization. As shown in Fig. \ref{TEDS}, in \emph{Stage I}, the focus is on reconstructing the  complete high-dimensional RIS received signal  based on the lower-dimensonal BS received signal,  while \emph{Stage II} employs the reconstructed RIS information to accurately localize the MU.

\subsection{Stage I: RIS Signal Reconstruction}
To enhance localization performance, it is imperative to restore the information lost due to the identified faulty elements. To this end, the primary objective of \emph{Stage I } is  the reconstruction of the complete high-dimensional signal received at the RIS.   As shown in the top half of Fig. \ref{TEDS}, \emph{Stage I} is based on an integrated approach using a  CNN  followed by a VAE \cite{sun2018learning}. Specifically, as shown in Fig. \ref{TEDS}, given ${\bm y}$, the  CNN is employed to  recover the signal from the non-faulty elements. Then, given the identified faulty elements and the signal from the non-faulty elements,  we restore the  signal lost by the faulty elements with the VAE.  We will first explain the rationale behind choosing the integrated CNN-VAE network, and subsequently outline the corresponding operational details. Finally, the integrated CNN-VAE network is described.
\subsubsection{Rationale for adapting CNN-VAE Network}
\paragraph{Rationale for using CNN}
CNNs are good at capturing complex function mappings and predicting continuous outputs, especially with a regression component \cite{CNNsurvey}. They effectively extract spatial features from signals using their Con layers \cite{Resnet}.
\paragraph{Rationale for using VAE}
VAE renowned for their adept data reconstruction from deficient inputs, emerge as a viable solution for image restoration \cite{sun2018learning}. Applying to the  reconstruction of the high-dimensional RIS received signal,  VAE can   capitalize on spatial correlations and the information of non-faulty elements. This enable VAE to learn the latent distribution of the input incomplete RIS signals, subsequently  generating  the complete RIS received signal  in alignment with this distribution \cite{sun2018learning}.
\paragraph{Rationale for Integrating CNN and VAE}
The integration of the CNN and VAE models provides significant benefits. Training them jointly allows shared feature extraction, improving model performance. This approach saves computational resources by avoiding multiple passes in separate networks and uses a single loss function for consistent training, leading to better signal reconstruction.
\subsubsection{Recovering Signal Recieved  from Non-Faulty Elements}
To reproduce the incomplete received signal at the RIS with faulty elements, as illustrated in Fig. \ref{TEDS}, a combination of a CNN and a regression module is employed, where the CNN is specifically utilized to estimate the signal, denoted as $ {\bm y}_r$. However, due to the   architecture of the CNN, its output cannot contain zero values. This leads to a potential misalignment between the CNN's output and the expected incomplete received signal. To overcome this challenge, we employ status vector $\bm{\mathcal{B}}$, where each element of the vector  is either $0$  or $1$. By multiplying the CNN's output with this status vector, the incomplete received signal at the RIS is successfully reproduced.
\paragraph{CNN Architecture}
As depicted in Fig. \ref{TEDS}, the proposed CNN has two Convolution Modules followed by an Output Module. Each Convolution Module encompasses a Con layer, a  ReLU activation function, and a max-pooling operation. The final Output Module consists of a linear layer with dimension $N_1 \times N_2 $.

\paragraph{CNN Training}
During   training, the primary goal is to condition the CNN, using received signal $ {\bm y} $ from the BS, to reproduce the received signal at the RIS, denoted as $ {\bm y}_r $. Besides, let us define the output of the employed CNN as $\hat{\bm y}_r$, which is given by
\begin{equation}
\hat{\bm y}_r= F_{\text{CNN}}({\bm y}; {\bm\theta }_{cnn})\odot \bm{\mathcal{B}},
\end{equation}
where  $ {\bm\theta }_{cnn} $ denotes the set of the parameters (or weights) of the CNN, and $F_{\text{CNN}}$ represents  a function mapping ${\bm y}$ to $\hat{\bm y}_r$. The objective of this training is to minimize the error caused by the reproduced signal as follows
\begin{equation}
\min_{{\bm\theta }_{cnn}} \ \lVert {\bm y}_r - \hat{\bm y}_r\rVert^2_2.
\end{equation}
Upon optimizing this objective, the CNN is conditioned to best recover the incomplete received signals at the RIS panel.

\subsubsection{Restoring Lost Information from Faulty Elements}
After recovering the incomplete RIS received signals through the CNN,  $\hat{\bm y}_r$ undergoes further reconstruction using a VAE as shown in Fig. \ref{TEDS}.

\paragraph{VAE Network Architecture}
According to Fig. \ref{TEDS}, the VAE begins with an encoder that learns the distribution of the input  $\hat{\bm y}_r $ through a series of linear layers. This process results in the establishment of a latent space, which represents the essential features of the data by determining the latent mean $\mu$ and latent variance $\sigma^2$. From this latent space, the encoder samples a set of latent variables. The decoder, which consists of another sequence of linear layers, uses these sampled latent variables to reconstruct the output, denoted as $\bar{\bm y}_r$ , which is an approximation of the input   \(\hat{\mathbf{y}}_r\) intended to be as close to the original input as possible. To be specific, the decoder randomly generates new instances within the latent space that are used to reconstruct the input data.

\paragraph{Training of VAE}
With the VAE structure in place, the reconstructed signal $ \bar{\bm y}_r $ can be obtained by feeding $ \hat{\bm y}_r $ through the encoder to get the latent variables, and subsequently passing these variables through the decoder. The specific operations are shown in Fig. \ref{TEDS}.   Mathematically, the reconstruction process can be represented as follows
\begin{align}
\bar{\bm y}_r = F_{\text{VAE}}(\hat{\bm y}_r; {\bm\theta }_{vae}),
\end{align}
where $ {\bm\theta }_{vae}$ denotes the  set of parameters (or weights) of the VAE, and $ F_{\text{VAE}}$ represents a function  mapping  $\hat{\bm y}_r$ to $\bar{\bm y}_r $. Hence, the  optimization problem for this training is formulated as
\begin{equation}
\min_{{\bm\theta }_{vae}} \ \lVert \hat{\bm y}_r - \bar{\bm y}_r \rVert^2_2+ \text{KL}(L_v),
\end{equation}
 where   KL and $L_v$ denote the  Kullback-Leibler divergence and the latent variables \cite{sun2018learning}. The inclusion of the KL-term acts as a regularization mechanism, ensuring that the latent variables adhere to a certain distribution.
Through this VAE-based reconstruction process, we aim to achieve a more precise and noise-resistant reconstruction of the RIS received signal.
\subsubsection{Integrated CNN-VAE Network}
The proposed integrated network combines the capabilities of CNN and VAE. The CNN is responsible for recovering the incomplete RIS received signal. Following this, the VAE reconstructs the complete RIS received signal.

\paragraph{Training}
During   training, the model is conditioned using the low-dimensional BS signal, $ {\bm y}$, to reconstruct complete high-dimensional RIS signal, $ \bar{\bm y}_r$. Simultaneously, the network incorporates the statuses of the faulty elements, $ \bm{\hat{\mathcal{B}}} $, to guide the learning process.
The objective during training is defined as
\begin{equation}
\begin{aligned}
\min_{{\bm\theta }_{cnn}, {\bm\theta }_{vae}} \Bigg[  \lVert {\bm y}_r - \hat{\bm y}_r\rVert^2_2+\lVert \hat{\bm y}_r - \bar{\bm y}_r \rVert^2_2  + \text{KL}(L_v) \Bigg].
\end{aligned}
\end{equation}

\paragraph{Output}
Upon convergence of  training, the network's parameters $ {\bm\theta }_{cnn}, {\bm\theta }_{vae}$ are optimized to  reconstruct the  RIS received signal. The model can then process any low-dimensional input $ {\bm y} $ to generate a higher-dimensional output $ \bar{\bm y}_r$.

\subsection{Stage II: Enhanced Localization using Reconstructed Signal $\bar{\bm y}_r$}
This stage builds upon the  reconstructed RIS information, $ \bar{\bm y}_r $, obtained in \emph{Stage I}. As shown in the lower half of Fig. \ref{TEDS}, the primary goal of this phase is to achieve enhanced localization using the reconstructed RIS information as well as the statuses of the elements. Similar to how we tackled the faulty element detection problem in the previous section, we propose to utilize the pre-trained DenseNet 121 for transfer learning to handle the localization problem described in \eqref{27}. By doing so, we can harness  the knowledge of the model acquired from previous tasks \cite{CNNsurvey}. This approach, grounded in the success of transfer learning for image classification \cite{transfer_learning}, is satisfied due to its smaller training times and often achieves better results comparing to the traditional training-from-scratch methods \cite{CNNsurvey}. Hence, we employ a transfer learning approach to localize the MU at \emph{Stage II}. As depicted in Fig. \ref{TEDS}, we input the reconstructed signal $ \bar{\bm y}_r $ as a type of fingerprint to the transferred DenseNet 121 to output the 3D location of the MU.

Before the training process, some modifications of the transferred DenseNet 121 are added to specifically address Problem (P2). The details are given as follows.
\subsubsection{Input Module}
To extract features from $\bar{\bm y}_r$, we divide it into its real and imaginary parts, represented as $\mathfrak{R}( \bar{\bm y}_r)$ and $\mathfrak{I}(\bar{\bm y}_r)$. These parts are then merged and reshaped based on the UPA antenna setup at the RIS. Since $N_1$ and $N_2$ may not fit the NN architecture's input size, we use interpolation techniques to make the signal dimensions compatible with DenseNet. Lastly, a Con layer adjusts the tensor size to meet the DenseNet 121 input requirements.
\begin{figure*}
    \centering
    \includegraphics[width=0.9\linewidth]{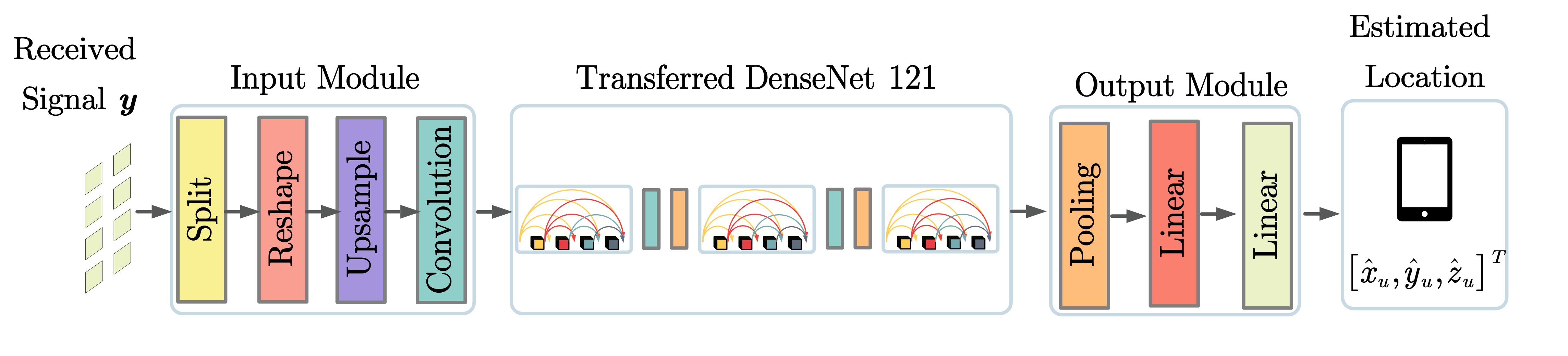}
    \caption{Network structure of  TEDF  algorithm.}
    \label{TEDF}
\end{figure*}
\subsubsection{Transferred DenseNet 121}
After preprocessing the received signal $\bar{\bm y}_r $, the  \emph{Input Module} outputs the preprocessed tensor to the transferred DenseNet 121. Notably, the internal layers of the transferred DenseNet 121 are initially frozen during the early training stage and gradually unfrozen as training progresses.
\subsubsection{Output Module}
Conventionally, DenseNet 121 employs a \emph{softmax} layer after the linear layer at the \emph{Output Module}  to yield category probabilities. To cater to the considered regression problem, we replace the \emph{softmax} layer with a \emph{linear} layer. This layer comprises three neurons, corresponding to the MU's coordinates.
\subsubsection{Selection of the Loss Function}
Predicting 3D coordinates is more effectively tackled using regression than classification, given that these coordinates exist in a continuous space rather than discrete categories \cite{3DCNN}. Trying to classify each coordinate as a unique category would drastically increase the number of classes, increasing complexity and computational demands \cite{Xiqi}. For regression tasks, the mean squared error (MSE) loss is often favored \cite{transfer_learning}. It measures the discrepancy between predicted and actual values and has continuous derivatives that facilitate optimization. Furthermore, larger errors are penalized more by MSE, which makes the model highly sensitive to significant deviations. Given the importance of precision in 3D coordinate prediction, regression with MSE emerges as a sound and efficient option.
\subsubsection{Transfer Learning for Localization}
Leveraging the foundational weights and configurations from pre-trained models, the focus shifts to training the DenseNet 121 for the task of localization. At this stage, the primary goal is to solve Problem (P2) which is mathematically represented as
\begin{equation}
\min_{\boldsymbol{\theta}_{\textrm{loc}}} \mathcal{L}_{\bm{p}}\left(\boldsymbol{\theta}_{\textrm{loc}}\right),
\end{equation}
where $\boldsymbol{\theta}_{\textrm{loc}}$ stands for the parameters (or weights) specific to the localization problem in the  transferred DenseNet 121.
\subsubsection{Localization with well-trained DenseNet}
Upon convergence of the training phase, the DenseNet 121 model stands ready to provide estimates for the 3D position of the MU. As we can see from Fig. \ref{TEDS}, for every reconstructed RIS singal $\bar{\bm y}_r$, the network outputs a three-dimensional vector, $[\hat{x}_u, \hat{y}_u, \hat{z}_u]^T$, which is an estimate of the MU's coordinates. Therefore, Problem (P2) has been effectively solved by the proposed TEDS algorithm.

\section{Transfer-Enhanced Direct Fingerprint Algorithm}\label{TEDFA}
To be able to show more light on the gain achieved by using RIS information and the impact of faulty RIS elements on localization, in this section, we propose a benchmark algorithm, named  TEDF algorithm, which only exploits the BS signal. As depicted in Fig. \ref{TEDF}, we directly input the BS received signal ${\bm y}$ as a type of fingerprint to the transferred DenseNet 121 to predict the 3D location of the MU.  In contrast to TEDS algorithm, this localization algorithm, without relying on RIS information, may suffer from some performance loss. Therefore, the comparison between these two kinds of algorithm can further highlight the necessity of faulty element detection and the benefits  of exploiting the high-dimensional RIS signal for localization.   The details of this algorithm are provided in the following.

\subsection{Transfer Learning for Localization}
The TEDF algorithm is also based on the DenseNet 121 model, leveraging its powerful transfer learning capabilities. However, in contrast to the method discussed in the previous section, the input data to this model is the received signal ${\bm y}$ at the BS. The fundamental rationale remains unchanged; the DenseNet is capable of deriving intricate spatial relationships from the signal, providing insights into the MU's position.
\subsection{Training Using BS Received Signal}
During the training phase,   the model learns to map the received signal ${\bm y}$ to the actual MU position without knowledge of the faulty elements. In a training process,   the model is optimized to reduce discrepancies between the predicted position and the actual MU position, using the same loss functions and methodologies as discussed earlier.
\subsection{Localization with Well-trained TEDF Algorithm}
After convergence of the training phase, the proposed TEDF algorithm  is capable of  estimating the 3D position of the MU. For every received  BS singal $ {\bm y} $, the network outputs a three-dimensional vector, $[\hat{x}_u, \hat{y}_u, \hat{z}_u]^T$, which corresponds to the MU coordinates.
\section{Simulation Results}

In this simulation, we consider a setup in a three-dimensional space, involving a MU, an RIS, and a BS. The carrier frequency is $f=90$ GHz. The RIS is composed of $ N_1 \times N_2 = 81 $ elements, with adjacent elements spaced at $d_r=\lambda/2=1.67\times10^{-3}$ m. The center of the RIS is located at $\bm{p}_r=(15, 0, 2)$ m. The BS comprises $ M_1 \times M_2 = 16 $ antennas, with an antenna spacing of $d_b=\lambda/2=1.67\times10^{-3}$ m. The center of the BS is positioned at $\bm{p}_b=(0, 10, 1.5) $ m.
The MU-RIS link includes $ P = 10 $ propagation paths, and the RIS-BS link comprises $ J = 10 $ propagation paths. The transmit power of the MU is $10$ dBm. The phase shifts of the elements at the RIS are set to a unity matrix.
\begin{figure*}[t]
    \centering
    \begin{minipage}{0.47\textwidth}
        \centering
        \includegraphics[width=\linewidth]{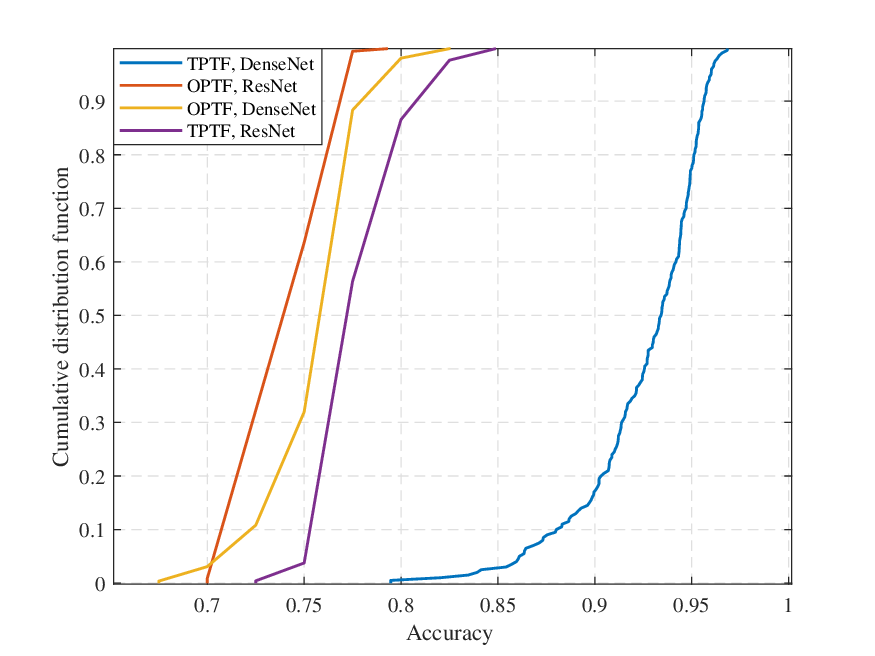}
        \caption{The CDF of detection accuracy with different kinds of algorithm.}\label{acc_alg}
    \end{minipage}%
    \hfill
    \begin{minipage}{0.47\textwidth}
        \centering
        \includegraphics[width=\linewidth]{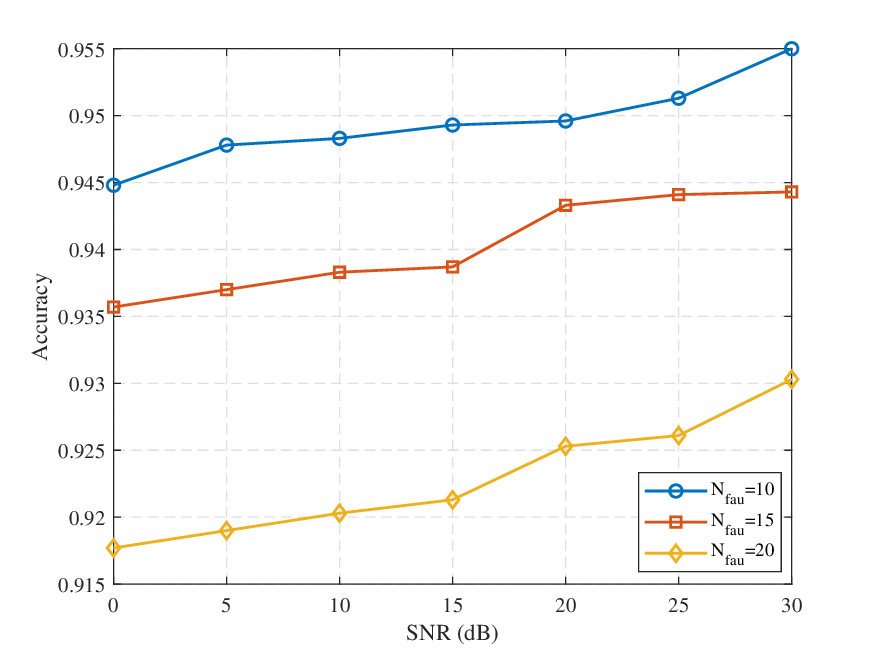}
        \caption{Detection accuracy versus SNR (dB) across different $N_\textrm{fau}$.}\label{acc_num}
    \end{minipage}
\end{figure*}

During the training stage of the proposed TPTL algorithm for detecting the faulty elements of the RIS, we assume the MU is located at $\bm{p}_u=(30,6,0.5)$ m, and the number of faulty elements within the RIS is assumed to be no more than $15$, i.e., $N_\textrm{faul}\leq15$. {
Furthermore, we assume that there are $K=9$ SAs, each containing $N_s = 9$ reflecting elements. The dataset consists of $20,000$ randomly generated fault scenarios, split into training and testing sets at a ratio of $0.8:0.2$.} During the training stage of the proposed TEDF and TEDS algorithms for estimating the location of the MU, we assume that the MU is uniformly distributed within three grids measuring $10$ m in length, $10$ m in width, and heights of $0.5$ m, $1.5$ m, and $2$ m, respectively. {The dataset consists of $60,000$ samples, each sample has the MU's coordinates, the condition of faulty elements on the RIS panel, and the corresponding signals received by both the BS and RIS. The dataset is split into training and testing sets at a ratio of $0.8:0.2$.}

Fig. \ref{acc_alg} presents the cumulative distribution function (CDF) of faulty element detection accuracy \footnote{{ The accuracy metric in our study requires identifying all faulty elements for a correct detection. The dataset consists of $20,000$ randomly generated fault scenarios, split into training and testing sets at a ratio of $0.8:0.2$.}} for different algorithms. The blue curve, labeled `TPTF DenseNet', signifies the TPTF algorithm transferring the DenseNet 121 architecture. Conversely, the purple curve, labeled `TPTF ResNet', represents the TPTF algorithm using ResNet 18. Furthermore, the yellow and red curves, labeled `OPTF DenseNet' and `OPTF ResNet' respectively, represent the benchmarks of the one phase transfer learning (OPTF) algorithm transferring DenseNet 121 and ResNet 18. As depicted in Fig. \ref{acc_alg}, the proposed TPTF, by sequentially detecting the faulty SAs and internal faulty elements, optimally harnesses DenseNet 121 or ResNet 18 for targeted faulty element detection, thus amplifying accuracy. Contrarily, the OPTF algorithm does not first detect the subarrays and then the elements within the subarrays, resulting in the number of elements (or labels) to be classified surpassing the performance capacity of DenseNet (or ResNet). Consequently, the performance of the OPTF algorithm is inferior. Significantly, the proposed `TPTF DenseNet' achieves commendable detection outcomes, with a $80\%$ probability of attaining a $91\%$ detection accuracy,  surpassing `TPTF ResNet', `OPTF DenseNet', and `OPTF ResNet', which have the accuracies of $76\%$, $74\%$, and $71\%$ at the identical confidence interval. In summary, Fig. \ref{acc_alg} emphasizes the superior detection performance of the propose `TPTF DenseNet' algorithm.
\begin{figure}[t]
    \centering
        \centering
        \includegraphics[width= \linewidth]{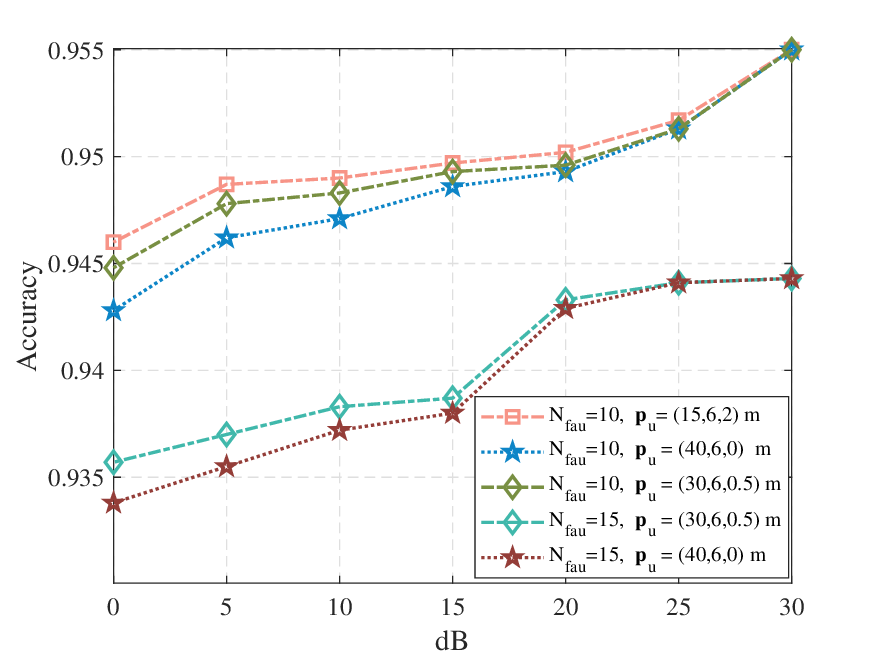}
        \caption{{Detection accuracy versus SNR (dB) across different positions and $N_\textrm{fau}$.}}\label{acc_position}
\end{figure}

{Fig. \ref{acc_num} presents the detection accuracy of the proposed TPTF algorithm under different signal-to-noise ratios (SNRs), showcasing its performance in test scenarios with a specified number of faulty elements, $N_\text{fau} = 10, 15, 20$. These numbers are exclusive to the testing phase.}  As delineated in Fig. \ref{acc_num}, the TPTF algorithm maintains remarkable efficacy, achieving a $92\%$ detection accuracy at an SNR of $10$ dB even  faced with $20$ faulty elements.   Moreover, in challenging conditions with an SNR as low as $0$ dB, the detection accuracies remain robust at approximately  $91.7\%$, $93.6\%$, and $94.5\%$ for $20$, $15$, and $10$ faulty elements, respectively. This underscores the robustness of the proposed TPTF detection algorithm in low SNR scenarios  \footnote{{ Exploring additional evaluation metrics such as precision, recall, and F-score represents an interesting avenue for future work, promising to provide deeper insights into model performance.}}.

{In Fig. \ref{acc_position}, we delve into the TPTL algorithm's effectiveness in identifying faulty elements under variable MU locations, particularly focusing on \({\bm p}_u = (15, 6, 2)\) m and \({\bm p}_u = (40, 6, 0)\) m with up to 10 faulty elements, extending to \({\bm p}_u = (40, 6, 0)\) m for scenarios encompassing up to 15 faulty elements. It is observed that there exists a  marginal improvement as the distance between the MU and the RIS-BS system narrows, especially in environments characterized by lower SNR values. However, the detection accuracy remains the same when SNR is larger than $20$ dB. This reveals that changes in the MU's position do have some effect, but it's not significant. Moreover, as the SNR increases, this impact diminishes.}

The essence of these observations lies in their testament to the TPTL algorithm's versatile performance across diverse spatial arrangements and signal conditions. This demonstrates not just the algorithm's capability to adjust its performance based on the proximity and configuration of the MU relative to the BS and RIS, but also its robustness and reliability in a variety of operational environments. This versatility is paramount for ensuring the algorithm's applicability and efficacy in real-world deployments, where environmental and situational variability is the norm.

\begin{figure*}
    \centering
        \begin{subfigure}{.30\textwidth}
        \centering
        \includegraphics[width=1\linewidth]{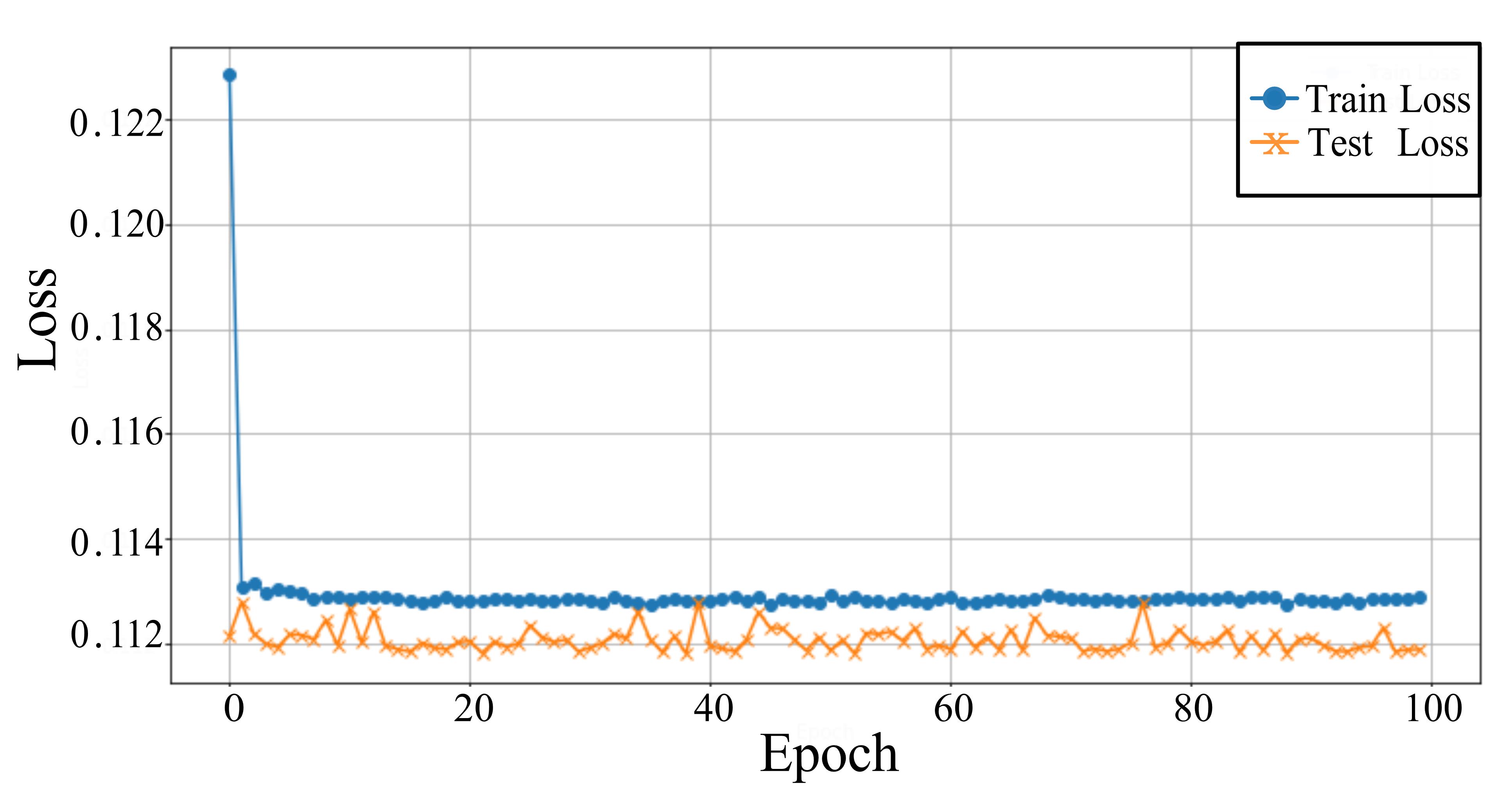}
        \caption{Test loss and train loss of  TEDF algorithm.}
        \label{TEDF_Loss}
    \end{subfigure}%
        \hfill
     \begin{subfigure}{.30\textwidth}
        \centering
        \includegraphics[width=1\linewidth]{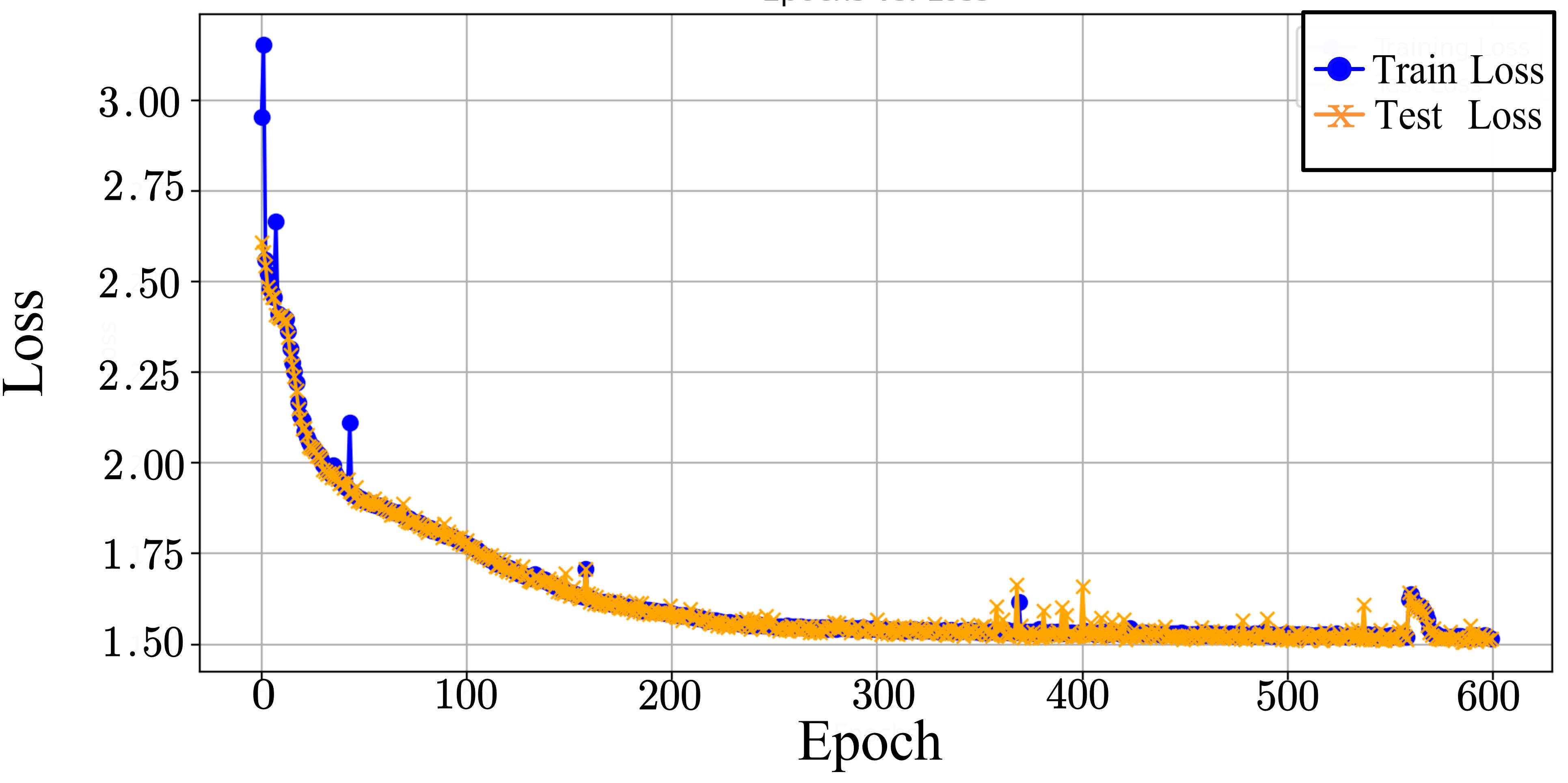}
        \caption{Test loss and train loss of \emph{Stage I} of  TEDS  algorithm.}
        \label{CNN_VAE_Loss}
    \end{subfigure}
    \hfill
    \begin{subfigure}{.30\textwidth}
        \centering
        \includegraphics[width=1\linewidth]{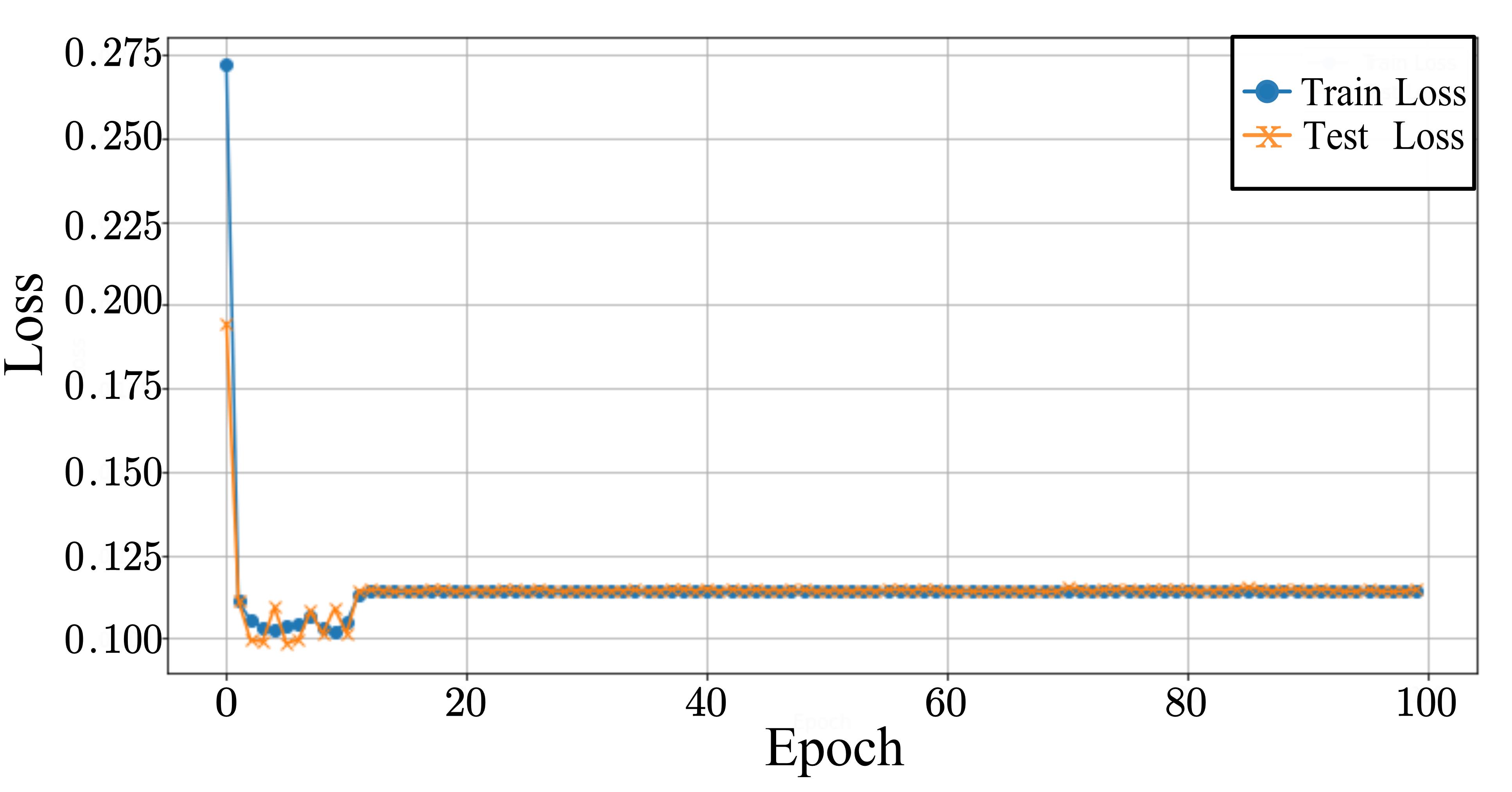}
        \caption{Test loss and train loss of \emph{Stage II} of  TEDS  algorithm.}
        \label{TEDS_Loss}
    \end{subfigure}%
 \hfill
    \caption{Losses of TEDF algorithm and TEDS algorithm in \emph{Stage I} and  \emph{Stage II}.}\label{Loss}
\end{figure*}
Fig. \ref{Loss} showcases the train and test loss curves of two  algorithms, i.e., TEDF and TEDS, employed for MU coordinate estimation. Specifically, Fig. \ref{TEDF_Loss} displays the iterative convergence behaviors of train and test  losses for the TEDF algorithm. It is evident that by approximately $20$ epochs, both train and test losses for the TEDF algorithm have converged. Although there are minor fluctuations in the test loss afterwards, their magnitude is negligible. Additionally, in \emph{Stage I} of  the TEDS algorithm, we employs the CNN-VAE network, which uses low-dimensional signals, ${\bm y}$,  to reconstruct complete high-dimensional RIS received signal at the RIS, $\bar{\bm y}_r$. Fig. \ref{CNN_VAE_Loss} presents the train and test losses of the proposed CNN-VAE network. During training, there is a rapid decline in both train and test losses initially, followed by a more gradual reduction. This demonstrates the complexity of reconstructing higher-dimensional information from low-dimensional information as well as the challenge of restoring signals expected to be received by faulty elements. The losses appear to stabilize around the $500$-th iteration. Lastly, Fig. \ref{TEDS_Loss} represents the convergence behavior for train and test  loss in \emph{Stage II} of the TEDS algorithm, where losses converge around the $15$-th epoch with negligible fluctuations. Comparing Fig. \ref{TEDS_Loss} and Fig. \ref{TEDF_Loss} which employ transfer learning with Fig. \ref{CNN_VAE_Loss}, the advantages of transfer learning are evident. By leveraging transfer learning, the number of required train epoches is significantly reduced while maintaining accuracy, making it highly suitable for the complex and dynamic environments anticipated in 6G. This offers valuable insights for future RIS-aided localization systems.

\begin{figure}[t]
    \centering
        \centering
        \includegraphics[width=0.8\linewidth]{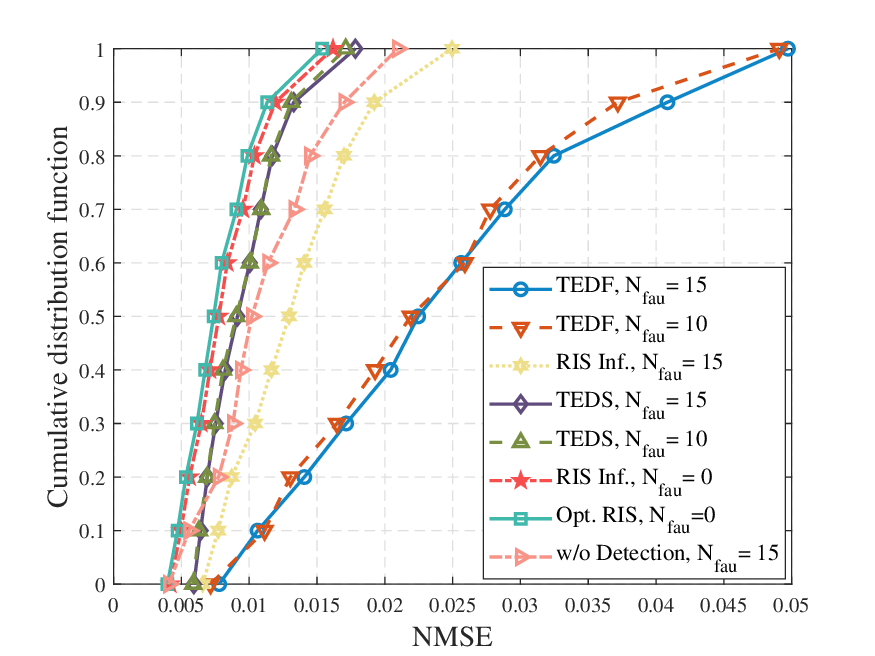}
        \caption{{Comparison of TEDF  and TEDS algorithms when $N_\textrm{fau}=10, 15$.}}\label{Finger_num}
\end{figure}

Fig. \ref{Finger_num} employs normalized mean square error (NMSE) to compare the localization accuracy of the TEDF and TEDS algorithms for $N_{\textrm{fau}}=10$ and $N_{\textrm{fau}}=15$. The curves labeled `TEDF, $N_{\textrm{fau}}=10$' and `TEDF, $N_{\textrm{fau}}=15$' correspond to scenarios using BS information as a fingerprint, while `TEDS, $N_{\textrm{fau}}=10$' and `TEDS, $N_\textrm{fau}=15$' represent the utilization of RIS information as a fingerprint for the same $N_{\textrm{fau}}$ values. The figure clearly shows that employing high-dimensional RIS information significantly improves localization performance compared to using lower-dimension BS information, even in the presence of faulty elements. This underscores the advantage of leveraging high-dimensional RIS information. Furthermore, the figure demonstrates that the localization performance remains relatively stable with minimal degradation when the number of faulty elements increases from $10$ to $15$, emphasizing the robustness of the proposed algorithms against faulty element variations.

Fig. \ref{Finger_num} also presents the curves representing scenarios where RIS information is used as a fingerprint but without processing through the CNN-VAE network. These scenarios are for RIS with no faulty elements and RIS containing 15 faulty elements,  labeled as`RIS Inf., $N_{\textrm{fau}}=0$', and `RIS Inf., $N_{\textrm{fau}}=15$', respectively. As a comparison in the picture,  the curves `TEDS, $N_{\textrm{fau}}=10$' and `TEDS, $N_{\textrm{fau}}=15$' show a minor gap with the `RIS Inf., $N_{\textrm{fau}}=0$' curve but a significant gap with the `RIS Inf., $N_{\textrm{fau}}=15$'curve.  This observation suggests the effectiveness of the proposed CNN-VAE network in reconstructing the complete high-dimensional RIS information and demonstrates that we can regain the localization accuracy with the proposed algorithms.

{Furthermore, Fig. \ref{Finger_num} also presents the scenarios that train the TEDS algorithm with the same loss but without the detection result, which is labeled as `w/o Detection, $N_\text{fau}= 15$'. It is shown that even though without the detection result of the faulty elements, the proposed integrated CNN-VAE network can still reconstruct the essential localization-related information.} Additionally, Fig. \ref{Finger_num} includes a curve labeled `Opt. RIS, $N_\textrm{fau}=0$', which represents scenarios where the information of a  RIS without faulty elements is used as a fingerprint, with its phase shifts optimized to maximize the received SNR \cite{Wuqq1}. As a comparison, there exists marginal gap between `Opt. RIS, $N_\textrm{fau}=0$' and `RIS Inf., $N_{\textrm{fau}}=0$', highlighting the robustness of the RIS information based localization algorithm  against the unoptimized phase shifts. We conjecture that the reason for this phenomenon lies in that the localization accuracy is dominated by the high-dimensional RIS information, and therefore the phase shift design does not play an important role.

\begin{figure}[t]
    \centering
        \centering
        \includegraphics[width=0.8\linewidth]{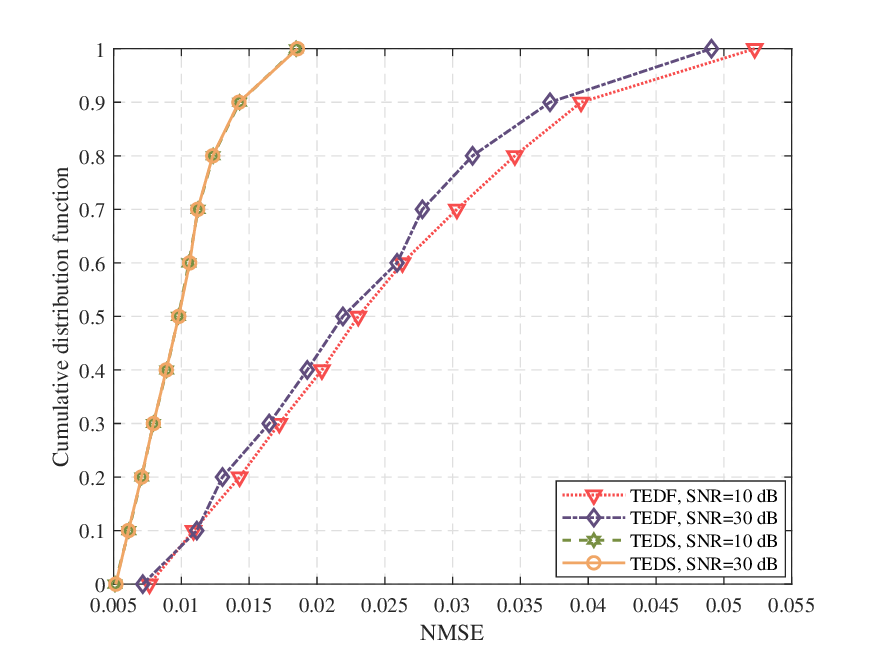}
        \caption{Comparison of TEDF and TEDS algorithms when $ \textrm{SNR}=10, 30$ dB.}\label{Finger_snr}
\end{figure}

Fig. \ref{Finger_snr} utilizes NMSE to compare the performance of the TEDF and TEDS algorithms under two distinct SNR conditions, i.e., $ \textrm{SNR}=10 $ dB and $ \textrm{SNR}=30$ dB. The red, purple, green, and orange curves respectively represent `TEDF, $ \textrm{SNR}=10 $ dB', `TEDF, $ \textrm{SNR}=30 $ dB', `TEDS, $ \textrm{SNR}=10 $ dB', and `TEDS, $ \textrm{SNR}=30 $ dB'. As depicted in Fig. \ref{Finger_snr}, the localization performance of the TEDF algorithm exhibits sensitivity to SNR variations, leading to a measurable reduction.   In contrast, the TEDS algorithm showcases remarkable robustness against SNR variations, as evidenced by the near overlap of the red and green curves.   We conjecture that the reason for this phenomenon lies in that  the employed CNN-VAE network can effectively against the variations of SNR. Such observations underscore the superiority of the proposed TEDS algorithm, suggesting that the localization algorithm remains highly effective even in diverse and intricate communication scenarios. This further supports the notion that localization performance can indeed benefit from the high-dimensional RIS information even with the faulty elements.

\begin{figure}[t]
    \centering
        \centering
        \includegraphics[width=0.8\linewidth]{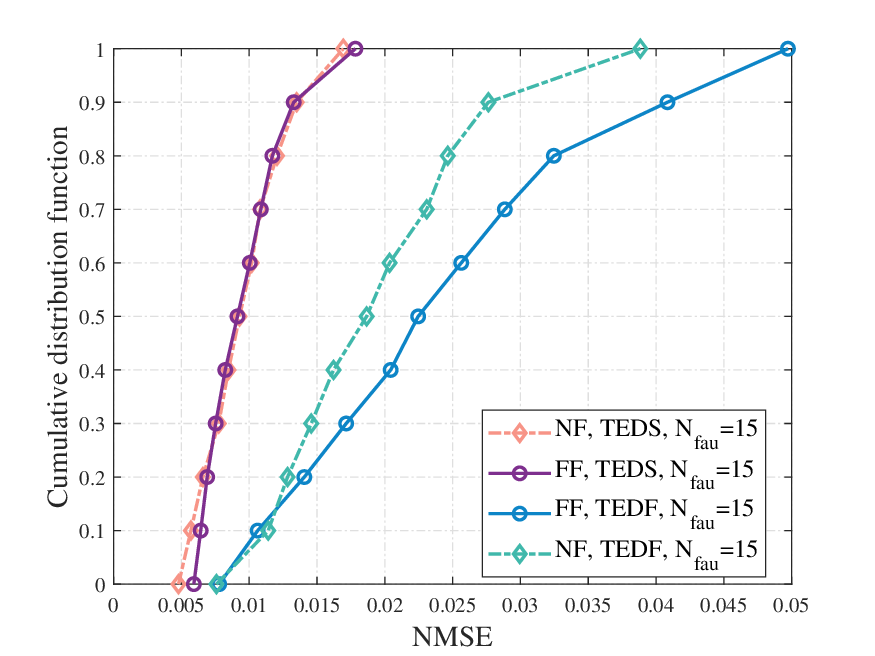}
        \caption{{NMSE Comparison between the far field and near field effect when employing the TEDS algorithm and TEDF algorithm}. }\label{NF_FF}
\end{figure}

{Fig. \ref{NF_FF} compares the NMSE of the proposed TEDS algorithm and TEDF algorithm with the near field and far field effects. As shown in Fig. \ref{NF_FF}, it can be observed that when employing the TEDF algorithm, the localization performance under near-field conditions significantly surpasses that in the far-field. However, with the application of the TEDS algorithm, while localization accuracy in the near-field still outperforms the far-field, the advantage is not as pronounced. We infer that this phenomenon is attributable to the richer localization information available in the near-field. Nonetheless, the dominant role of the high-dimensional RIS information reconstructed by the TEDS algorithm mitigates the disparity in gains between the near and far fields. Moreover, Fig. \ref{NF_FF}  also corroborates the effectiveness of the proposed algorithm in both near-field and far-field conditions, underscoring its robustness and adaptability across different propagation environments.}

\begin{figure}[t]
    \centering
        \centering
        \includegraphics[width=0.8\linewidth]{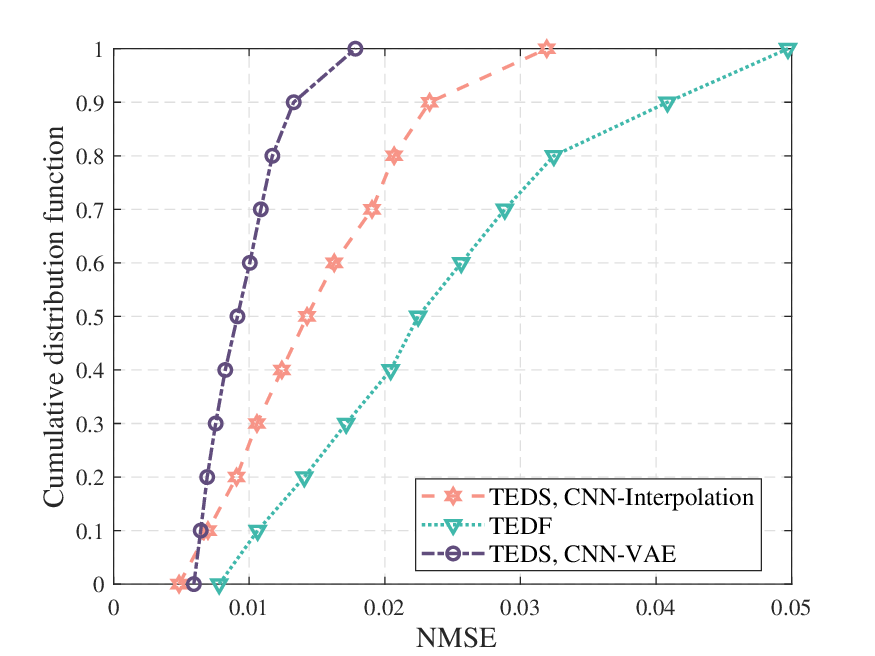}
        \caption{{NMSE Comparison between the Integrated CNN-VAE method and the CNN-Interpolation method}. }\label{CNN-no-VAE}
\end{figure}

{Fig. \ref{CNN-no-VAE} compares the NMSE of the proposed TEDS algorithm with the integrated CNN-VAE network and with the CNN-Interpolation method. As shown in Fig. \ref{CNN-no-VAE}, the CNN-Interpolation method shows some improvement over the TEDF algorithm. However, there is still a significant gap when compared to the TEDS algorithm that utilizes the CNN-VAE network. This comparison validates the superiority of our proposed TEDS algorithm, especially in addressing the capability of restoring the lost RIS information} \footnote{{
In this study, the VAE model is trained to reconstruct complete RIS information from the incomplete data represented by $\bar{\bm y}_r$. For future endeavors, an appealing research direction would be to supervise the training of the VAE with complete RIS information originating from an undamaged RIS, presenting a substantial avenue for further exploration and refinement. }}.

\section{Conclusion}\label{Con}
This paper has investigated  the  localization algorithm design based on the high-dimensional RIS information   in the presence of faulty RIS elements. We have designed a  TPTL algorithm for faulty element detection.  We have proposed  a TEDS  algorithm to clear their harmful impact, unleashing the gain of RIS on localization.   At \emph{Stage I}, we  have reconstructed the complete high-dimensional  RIS received signal from the low-dimensional  BS received signal. At \emph{Stage II}, this reconstructed RIS information  has been input to the transferred DenseNet  to estimate the location of the MU. We  have also proposed  a low-complexity TEDF algorithm which only utilizes the low-dimensional BS information. Simulation results have demonstrated that we can effectively benefit from the high-dimensional RIS information even with faulty elements.
\bibliographystyle{IEEEtran}
				\bibliography{myre}
				\begin{IEEEbiography}[{\includegraphics[width=0.8 in,clip,keepaspectratio]{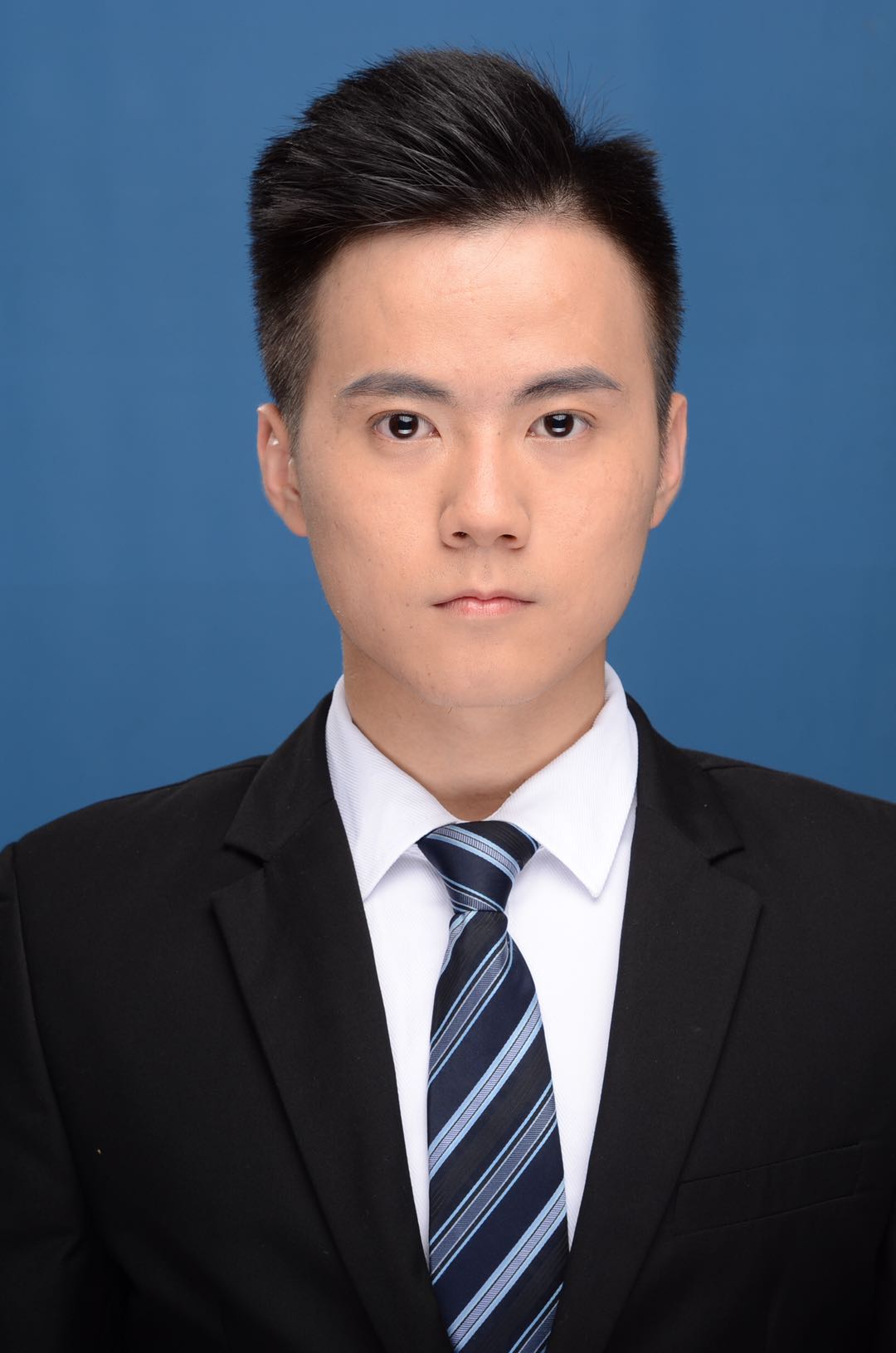}}]
	{Tuo Wu}  received the B.Eng. degree in telecommunication engineering from South China Normal University, Guangzhou, China, in 2017, and the M.S. degree in wireless radio physics from Sun Yat-Sen University, Guangzhou, China, in 2021. He is currently working toward the Ph.D. degree with the School of Electronic Engineering and Computer Science, Queen Mary University of London, London, U.K. His research interests include Reconfigurable Intelligent Surface and localization.
\end{IEEEbiography}
\begin{IEEEbiography}[{\includegraphics[width=1.0 in,clip,keepaspectratio]{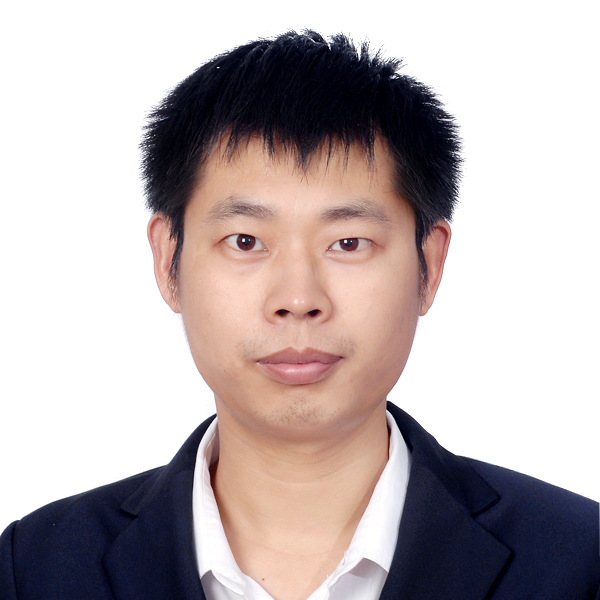}}]
{Cunhua Pan}
received the B.S. and Ph.D. degrees from the School of Information Science and Engineering, Southeast University, Nanjing, China, in 2010 and 2015, respectively. From 2015 to 2016, he was a Research Associate at the University of Kent, U.K. He held a post-doctoral position at Queen Mary University of London, U.K., from 2016 and 2019.From 2019 to 2021, he was a Lecturer in the same university. From 2021, he is a full professor in Southeast University.

His research interests mainly include reconfigurable intelligent surfaces (RIS), intelligent reflection surface (IRS), ultra-reliable low latency communication (URLLC) , machine learning, UAV, Internet of Things, and mobile edge computing. He has published over 120 IEEE journal papers. He is currently an Editor of IEEE Transactions on Vehicular Technology, IEEE Wireless Communication Letters, IEEE Communications Letters and IEEE ACCESS. He serves as the guest editor for IEEE Journal on Selected Areas in Communications on the special issue on xURLLC in 6G: Next Generation Ultra-Reliable and Low-Latency Communications. He also serves as a leading guest editor of IEEE Journal of Selected Topics in Signal Processing (JSTSP) Special Issue on Advanced Signal Processing for Reconfigurable Intelligent Surface-aided 6G Networks, leading guest editor of IEEE Vehicular Technology Magazine on the special issue on Backscatter and Reconfigurable Intelligent Surface Empowered Wireless Communications in 6G, leading guest editor of IEEE Open Journal of Vehicular Technology on the special issue of Reconfigurable Intelligent Surface Empowered Wireless Communications in 6G and Beyond, and leading guest editor of IEEE ACCESS Special Issue on Reconfigurable Intelligent Surface Aided Communications for 6G and Beyond. He is Workshop organizer in IEEE ICCC 2021 on the topic of Reconfigurable Intelligent Surfaces for Next Generation Wireless Communications (RIS for 6G Networks), and workshop organizer in IEEE Globecom 2021 on the topic of Reconfigurable Intelligent Surfaces for future wireless communications. He is currently the Workshops and Symposia officer for Reconfigurable Intelligent Surfaces Emerging Technology Initiative. He is workshop chair for IEEE WCNC 2024, and TPC co-chair for IEEE ICCT 2022. He serves as a TPC member for numerous conferences, such as ICC and GLOBECOM, and the Student Travel Grant Chair for ICC 2019. He received the IEEE ComSoc Leonard G. Abraham Prize in 2022, IEEE ComSoc Asia-Pacific Outstanding Young Researcher Award, 2022.
\end{IEEEbiography}
\begin{IEEEbiography}[{\includegraphics[width=1.0 in,clip,keepaspectratio]{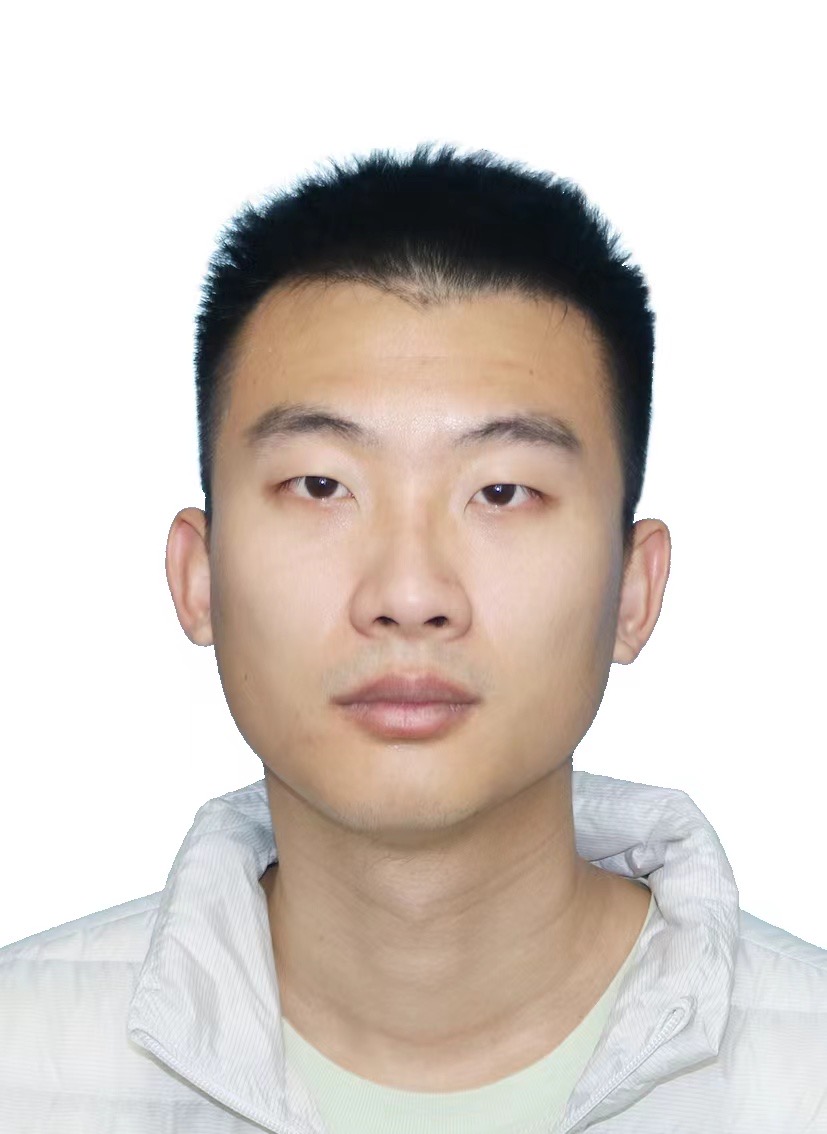}}]
	{Kangda Zhi}  received the B.Eng degree from the School of Communication and Information Engineering, Shanghai University, Shanghai, China, in 2017, the M.Eng degree from School of Information Science and Technology, University of Science and Technology of China, Hefei, China, in 2020, and the Ph.D. degree from the School of Electronic Engineering and Computer Science, Queen Mary University of London, U.K., in 2023. He is currently a Postdoctoral Researcher with the School of Electrical Engineering and Computer Science, Technical University of Berlin, Germany. His research interests include Reconfigurable Intelligent Surface, massive MIMO, and near-field communications. He received the Exemplary Reviewer Certificate of the IEEE WIRELESS COMMUNICATIONS LETTERS in 2021 and 2022 and the IEEE COMMUNICATIONS LETTERS in 2023.
\end{IEEEbiography}

\begin{IEEEbiography}[{\includegraphics[width=1.1 in,clip,keepaspectratio]{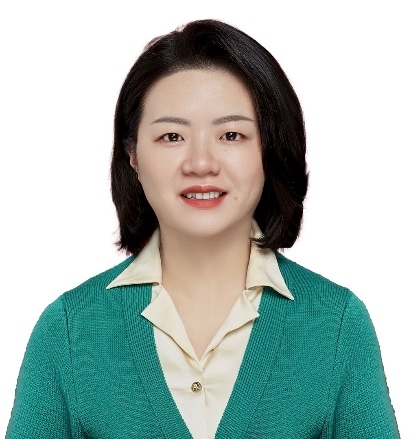}}]
	{Hong Ren} received the B.S. degree in electrical engineering from Southwest Jiaotong University, Chengdu, China, in 2011, and the M.S. and Ph.D. degrees in electrical engineering from Southeast University, Nanjing, China, in 2014 and 2018, respectively. From 2016 to 2018, she was a Visiting Student with the School of Electronics and Computer Science, University of Southampton, U.K. From 2018 to 2020, she was a Post-Doctoral Scholar with Queen Mary University of London, U.K. She is currently an associate professor with Southeast University. Her research interests lie in the areas of communication and signal processing, including ultra-low latency and high reliable communications, Massive MIMO and machine learning.
\end{IEEEbiography}

\begin{IEEEbiography}[{\includegraphics[width=1.0 in,clip,keepaspectratio]{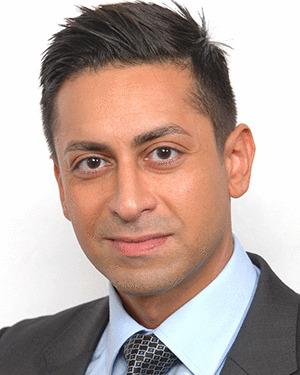}}]
	{Maged Elkashlan} received the PhD degree in Electrical Engineering from the University of British Columbia in 2006. From 2007 to 2011, he was a Scientist at the Commonwealth Scientific and Industrial Research Organization (CSIRO) Australia. During this time, he held visiting faculty appointments at University of New South Wales, University of Sydney, and University of Technology Sydney. In 2011, he joined the School of Electronic Engineering and Computer Science at Queen Mary University of London. He also holds a visiting faculty appointment at Beijing University of Posts and Telecommunications. His research interests fall into the broad areas of communication theory and statistical signal processing.

Dr. Elkashlan is an Editor of the IEEE TRANSACTIONS ON COMMUNICATIONS and the IEEE TRANSACTIONS ON VEHICULAR TECHNOLOGY. He served as an Editor of the IEEE TRANSACTIONS ON MOLECULAR, BIOLOGICAL AND MULTI-SCALE COMMUNICATIONS from 2015 to 2022, IEEE TRANSACTIONS ON WIRELESS COMMUNICATIONS from 2013 to 2018, and IEEE COMMUNICATIONS LETTERS from 2012 to 2016. He also served as a Guest Editor of the special issue on "Location Awareness for Radios and Networks" of IEEE JOURNAL ON SELECTED AREAS IN COMMUNICATIONS, the special issue on "Energy Harvesting Communications" of IEEE COMMUNICATIONS MAGAZINE, the special issue on "Green Media: The Future of Wireless Multimedia Networks" of IEEE WIRELESS COMMUNICATIONS, and the special issue on "Millimeter Wave Communications for 5G" of IEEE COMMUNICATIONS MAGAZINE.
\end{IEEEbiography}

\begin{IEEEbiography}[{\includegraphics[width=1.0 in,clip,keepaspectratio]{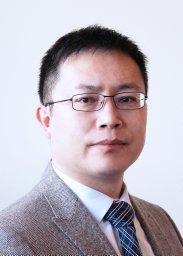}}]
	{Cheng-Xiang Wang}  (Fellow, IEEE) received the B.Sc. and M.Eng. degrees in communication and information systems from Shandong University, China, in 1997 and 2000, respectively, and the Ph.D. degree in wireless communications from Aalborg University, Denmark, in 2004.

He was a Research Assistant with the Hamburg University of Technology, Hamburg, Germany, from 2000 to 2001, a Visiting Researcher with Siemens AG Mobile Phones, Munich, Germany, in 2004, and a Research Fellow with the University of Agder, Grimstad, Norway, from 2001 to 2005. He was with Heriot-Watt University, Edinburgh, U.K., from 2005 to 2018, where he was promoted to a professor in 2011. He has been with Southeast University, Nanjing, China, as a professor since 2018, and he is now the Executive Dean of the School of Information Science and Engineering. He is also a professor with Pervasive Communication Research Center, Purple Mountain Laboratories, Nanjing, China. He has authored 4 books, 3 book chapters, and over 550 papers in refereed journals and conference proceedings, including 27 highly cited papers. He has also delivered 27 invited keynote speeches/talks and 18 tutorials in international conferences. His current research interests include wireless channel measurements and modeling, 6G wireless communication networks, and electromagnetic information theory.

Dr. Wang is a Member of the Academia Europaea (The Academy of Europe), a Member of the European Academy of Sciences and Arts (EASA), a Fellow of the Royal Society of Edinburgh (FRSE), IEEE, IET and China Institute of Communications (CIC), an IEEE Communications Society Distinguished Lecturer in 2019 and 2020, a Highly-Cited Researcher recognized by Clarivate Analytics in 2017-2020. He is currently an Executive Editorial Committee Member of the IEEE TRANSACTIONS ON WIRELESS COMMUNICATIONS. He has served as an Editor for over ten international journals, including the IEEE TRANSACTIONS ON WIRELESS COMMUNICATIONS, from 2007 to 2009, the IEEE TRANSACTIONS ON VEHICULAR TECHNOLOGY, from 2011 to 2017, and the IEEE TRANSACTIONS ON COMMUNICATIONS, from 2015 to 2017. He was a Guest Editor of the IEEE JOURNAL ON SELECTED AREAS IN COMMUNICATIONS, Special Issue on Vehicular Communications and Networks (Lead Guest Editor), Special Issue on Spectrum and Energy Efficient Design of Wireless Communication Networks, and Special Issue on Airborne Communication Networks. He was also a Guest Editor for the IEEE TRANSACTIONS ON BIG DATA, Special Issue on Wireless Big Data, and is a Guest Editor for the IEEE TRANSACTIONS ON COGNITIVE COMMUNICATIONS AND NETWORKING, Special Issue on Intelligent Resource Management for 5G and Beyond. He has served as a TPC Member, a TPC Chair, and a General Chair for more than 30 international conferences. He received 16 Best Paper Awards from IEEE GLOBECOM 2010, IEEE ICCT 2011, ITST 2012, IEEE VTC 2013 Spring, IWCMC 2015, IWCMC 2016, IEEE/CIC ICCC 2016, WPMC 2016, WOCC 2019, IWCMC 2020, WCSP 2020, CSPS2021, WCSP 2021, IEEE/CIC ICCC 2022, and IEEE ICCT 2023.
\end{IEEEbiography} 	

\begin{IEEEbiography}[{\includegraphics[width=1.3 in, angle=90, clip,keepaspectratio]{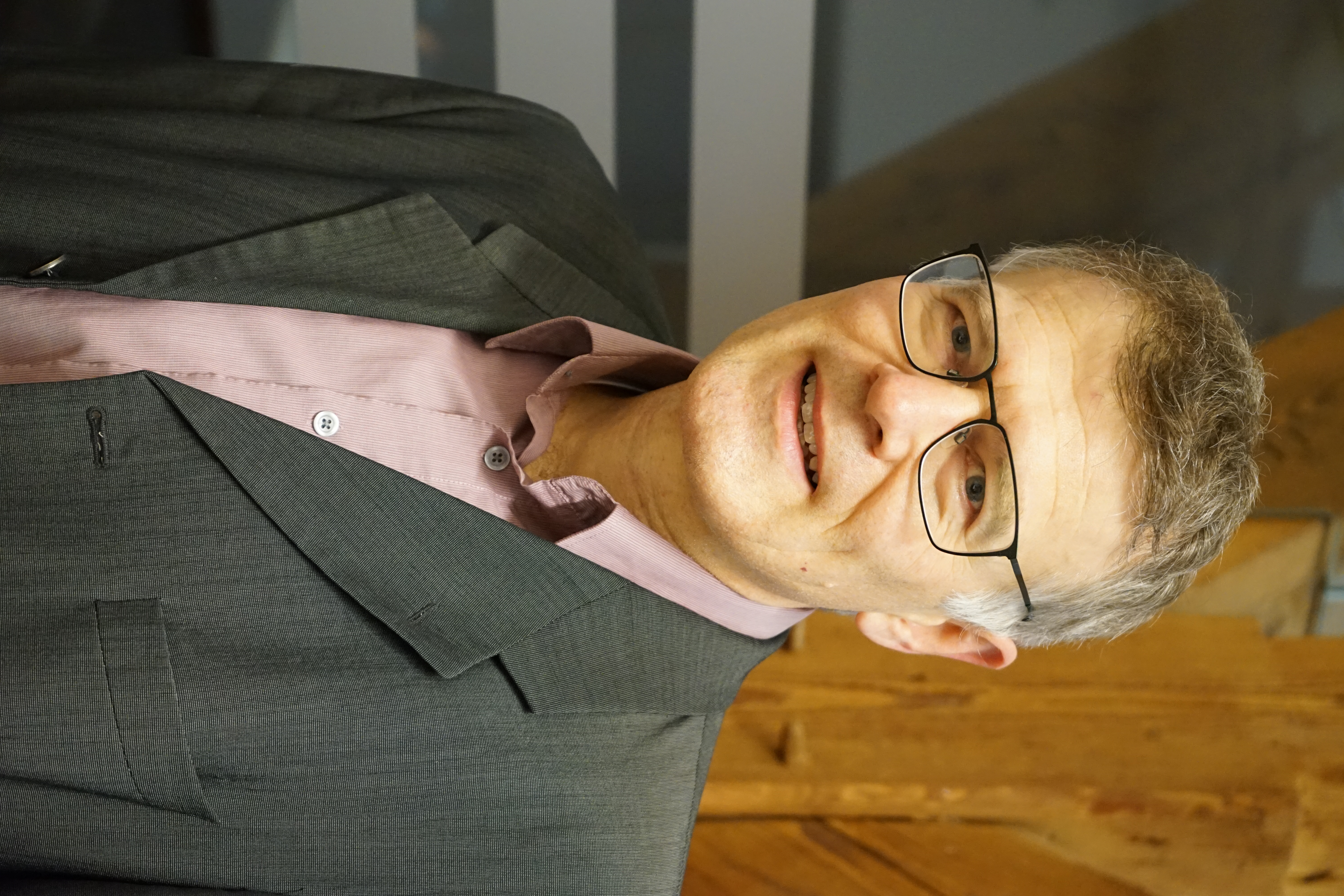}}]
	{Robert Schober } (S'98, M'01, SM'08, F'10) received the Diplom (Univ.) and the Ph.D. degrees in electrical engineering from Friedrich-Alexander University of Erlangen-Nuremberg (FAU), Germany, in 1997 and 2000, respectively. From 2002 to 2011, he was a Professor and Canada Research Chair at the University of British Columbia (UBC), Vancouver, Canada. Since January 2012 he is an Alexander von Humboldt Professor and the Chair for Digital Communication at FAU. His research interests fall into the broad areas of Communication Theory, Wireless and Molecular Communications, and Statistical Signal Processing.

Robert received several awards for his work including the 2002 Heinz Maier Leibnitz Award of the German Science Foundation (DFG), the 2004 Innovations Award of the Vodafone Foundation for Research in Mobile Communications, a 2006 UBC Killam Research Prize, a 2007 Wilhelm Friedrich Bessel Research Award of the Alexander von Humboldt Foundation, the 2008 Charles McDowell Award for Excellence in Research from UBC, a 2011 Alexander von Humboldt Professorship, a 2012 NSERC E.W.R. Stacie Fellowship, a 2017 Wireless Communications Recognition Award by the IEEE Wireless Communications Technical Committee, and the 2022 IEEE Vehicular Technology Society Stuart F. Meyer Memorial Award. Furthermore, he received numerous Best Paper Awards for his work including the 2022 ComSoc Stephen O. Rice Prize and the 2023 ComSoc Leonard G. Abraham Prize. Since 2017, he has been listed as a Highly Cited Researcher by the Web of Science. Robert is a Fellow of the Canadian Academy of Engineering, a Fellow of the Engineering Institute of Canada, and a Member of the German National Academy of Science and Engineering.

He served as Editor in Chief of the IEEE Transactions on Communications, VP Publications of the IEEE Communication Society (ComSoc), ComSoc Member at Large, and ComSoc Treasurer. Currently, he serves as Senior Editor of the Proceedings of the IEEE and as ComSoc President.	
\end{IEEEbiography} 			

\begin{IEEEbiography}[{\includegraphics[width=1.0 in,   clip,keepaspectratio]{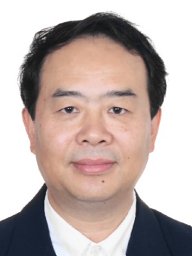}}]
	{Xiao-Hu You} (IEEE Fellow) received his Master and Ph.D. Degrees from Southeast University, Nanjing, China, in
Electrical Engineering in 1985 and 1988, respectively. Since 1990, he has been working with National Mobile Communications Research Laboratory at Southeast University, where he is currently professor and director of the Lab. He has contributed over 300 IEEE journal papers and 3 books in the areas of wireless communications. From 1999 to 2002, he was the Principal Expert of China C3G Project. From 2001-2006, he was the Principal Expert of
China National 863 Beyond 3G FuTURE Project. From 2013 to 2019, he was the Principal Investigator of China National 863 5G Project. His current research interests include wireless networks, advanced signal processing and its applications.

Dr. You was selected as IEEE Fellow in 2011. He served as the General Chair for IEEE Wireless Communications and Networking Conference (WCNC) 2013, IEEE Vehicular Technology Conference (VTC) 2016 Spring,
and IEEE International Conference on Communications (ICC) 2019.
\end{IEEEbiography} 		
			\end{document}